\documentclass[prd,aps,floats,floatfix,eqsecnum,nofootinbib]{revtex4}
\usepackage{graphics}
\usepackage{graphicx}
\usepackage{rotate}
\usepackage{color}
\usepackage{rotating}
\usepackage{psfrag}
\usepackage{amsmath,amsthm}
\usepackage{epsfig}
\usepackage{slashed}
\usepackage{sidecap}

\begin{document}
\newcommand{\D}{\displaystyle} 
\newcommand{\T}{\textstyle} 
\newcommand{\SC}{\scriptstyle} 
\newcommand{\SSC}{\scriptscriptstyle} 
\newcommand{\be}{\begin{equation}}
\newcommand{\ee}{\end{equation}}
\newcommand{\avg}[1]{\langle #1 \rangle}
\newcommand{\vx}{{\boldsymbol{x}}}
\newcommand{\rv}{{\boldsymbol{r}}}
\newcommand{\vq}{\ensuremath{\vec{q}}}
\newcommand{\pv}{\ensuremath{\vec{p}}}
\def\AJ{{\it Astron. J.} }
\def\ARAA{{\it Annual Rev. of Astron. \& Astrophys.} }
\def\ApJ{{\it Astrophys. J.} }
\def\ApJL{{\it Astrophys. J. Letters} }
\def\ApJS{{\it Astrophys. J. Suppl.} }
\def\ApP{{\it Astropart. Phys.} }
\def\AA{{\it Astron. \& Astroph.} }
\def\AAR{{\it Astron. \& Astroph. Rev.} }
\def\AAL{{\it Astron. \& Astroph. Letters} }
\def\AASu{{\it Astron. \& Astroph. Suppl.} }
\def\AN{{\it Astron. Nachr.} }
\def\IJMP{{\it Int. J. of Mod. Phys.} }
\def\JGR{{\it Journ. of Geophys. Res.}}
\def\JHEP{{\it Journ. of High En. Phys.} }
\def\JPhG{{\it Journ. of Physics} {\bf G} }
\def\MNRAS{{\it Month. Not. Roy. Astr. Soc.} }
\def\Nature{{\it Nature} }
\def\NewAR{{\it New Astron. Rev.} }
\def\NJPh{{\it New Journ. of Phys.} }
\def\PASP{{\it Publ. Astron. Soc. Pac.} }
\def\PhFl{{\it Phys. of Fluids} }
\def\PLB{{\it Phys. Lett.}{\bf B} }
\def\PhysRep{{\it Phys. Rep.} }
\def\PR{{\it Phys. Rev.} }
\def\PRD{{\it Phys. Rev.} {\bf D} }
\def\PRL{{\it Phys. Rev. Letters} }
\def\RMP{{\it Rev. Mod. Phys.} }
\def\Science{{\it Science} }
\def\ZfA{{\it Zeitschr. f{\"u}r Astrophys.} }
\def\ZfN{{\it Zeitschr. f{\"u}r Naturforsch.} }
\def\etal{{\it et al.}}
\hyphenation{mono-chro-matic sour-ces Wein-berg
chang-es Strah-lung dis-tri-bu-tion com-po-si-tion elec-tro-mag-ne-tic
ex-tra-galactic ap-prox-i-ma-tion nu-cle-o-syn-the-sis re-spec-tive-ly
su-per-nova su-per-novae su-per-nova-shocks con-vec-tive down-wards
es-ti-ma-ted frag-ments grav-i-ta-tion-al-ly el-e-ments me-di-um
ob-ser-va-tions tur-bul-ence sec-ond-ary in-ter-action
in-ter-stellar spall-ation ar-gu-ment de-pen-dence sig-nif-i-cant-ly
in-flu-enc-ed par-ti-cle sim-plic-i-ty nu-cle-ar smash-es iso-topes
in-ject-ed in-di-vid-u-al nor-mal-iza-tion lon-ger con-stant
sta-tion-ary sta-tion-ar-i-ty spec-trum pro-por-tion-al cos-mic
re-turn ob-ser-va-tion-al es-ti-mate switch-over grav-i-ta-tion-al
super-galactic com-po-nent com-po-nents prob-a-bly cos-mo-log-ical-ly
Kron-berg Berk-huij-sen}

\title{\Large HIGHLIGHTS and CONCLUSIONS 

\medskip

of the Chalonge CIAS Meudon Workshop 2010:

\medskip

`Dark Matter in the Universe and Universal Properties of Galaxies: 
Theory and Observations',    

\medskip

Ecole Internationale d'Astrophysique Daniel Chalonge

Meudon campus of Observatoire de Paris

in the historic Ch\^ateau, 8-11 June 2010.}

\author{\Large \bf   H.J. de Vega$^{(a,b)}$,    N.G. Sanchez$^{(b)}$}

\date{\today}

\affiliation{$^{(a)}$ LPTHE, Universit\'e
Pierre et Marie Curie (Paris VI) et Denis Diderot (Paris VII),
Laboratoire Associ\'e au CNRS UMR 7589, Tour 24, 5\`eme. \'etage, 
Boite 126, 4, Place Jussieu, 75252 Paris, Cedex 05, France. \\
$^{(b)}$ Observatoire de Paris,
LERMA. Laboratoire Associ\'e au CNRS UMR 8112.
 \\61, Avenue de l'Observatoire, 75014 Paris, France.}

\maketitle

\tableofcontents

\begin{figure}[htbp]
\epsfig{file=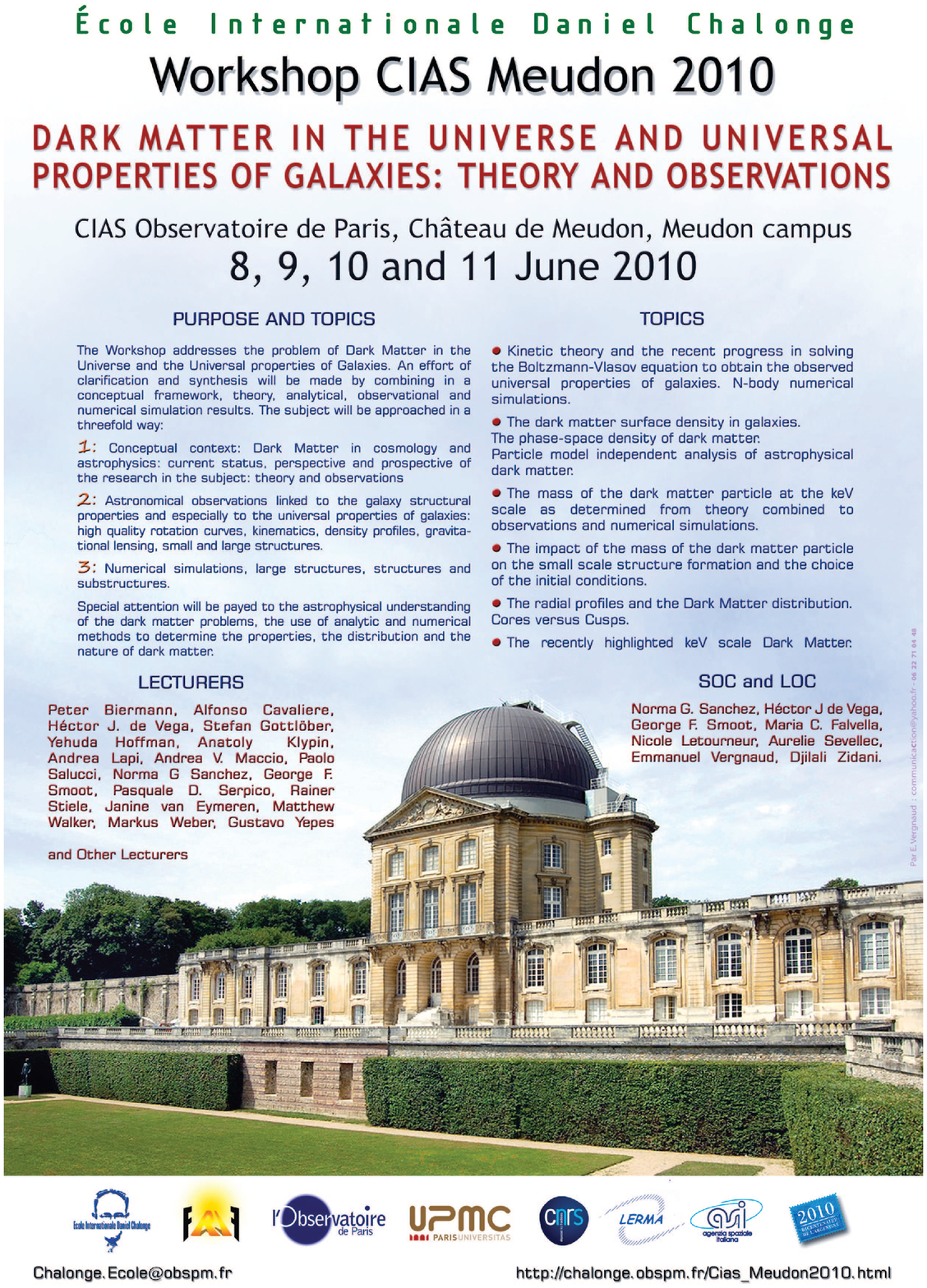,width=14cm,height=18cm}
\caption{Poster of the Workshop}
\end{figure}

\begin{figure}[h]
\includegraphics[scale=.7]{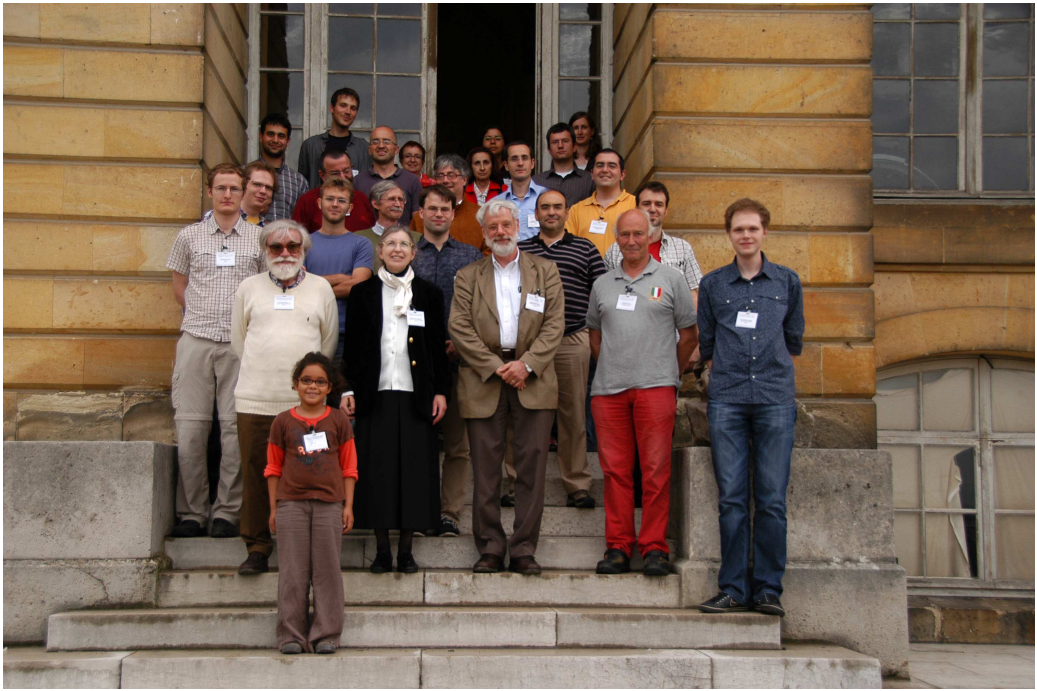}
\caption{Photo of the Group}
\end{figure}

\section{Purpose of the Workshop and Introduction}

The Workshop addressed the problem of Dark Matter in the 
Universe and the Universal 
properties of Galaxies, with an effort of clarification and synthesis by 
combining in a conceptual framework, theory, analytical, observational and 
numerical simulation results. 

The subject  have been approached in a threefold way:

(I) Conceptual context: Dark Matter in cosmology and astrophysics: 
current status, perspective and prospective of the research in 
the subject: theory and observations. 

(II) Astronomical observations linked to the 
galaxy structural properties and especially to the universal properties of 
galaxies: high quality rotation curves, kinematics, density profiles, 
gravitational lensing, small and large structures. 

(III) Numerical simulations, large structures, structures and substructures.  

\medskip

Topics addressed by the Workshop included:
Kinetic theory and the recent progress in solving the 
Boltzmann-Vlasov equation to obtain the observed universal 
properties of galaxies. 
The dark matter surface density in galaxies. 
The phase-space density of dark matter. 
Observations of Galaxy properties, surface density and universal profiles 
Discrepancies between numerical simulations results and 
observations.          
Particle model independent analysis of astrophysical dark matter. 
The mass of the dark matter particle at the keV scale from theory 
combined to 
observations and numerical simulations. The impact of the mass of the dark 
matter particle on the small scale structure formation and the 
choice of the initial 
conditions. The radial profiles and the Dark Matter distribution. 
Cores vs. Cusps. 
The recently highlighted keV scale Dark Matter.

\medskip

Sessions lasted for four full days in the beautiful Meudon campus 
of Observatoire de Paris, 
where CIAS `Centre International d'Ateliers Scientifiques' is located. 
All sessions took place in the historic Meudon Ch\^ateau, 
(built in 1706 by great architect Jules-Hardouin Mansart in orders by 
King Louis XIV for his son the Grand Dauphin). 

The Meeting was open to all scientists interested in the subject. 
All sessions were plenary followed by discussions. 
The format of the Meeting was intended to allow easy and 
fruitful mutual contact and communication with large time devoted  
to discussions. All Informations about the meeting, are displayed at

\begin{center}

{\bf http://www.chalonge.obspm.fr/Cias\_Meudon2010.html}

\end{center}

The presentations by the lecturers are available on line 
(in .pdf format) in `Programme and Lecturers' in the above link, 
as well as the photos of 
the Workshop. We thank all again, both lecturers and participants, 
for having contributed so much to the great success of this 
Workshop and look forward to seeing you again in the next 
Workshop of this series.     

We thank the Observatoire de Paris and the CIAS support, as well as the
logistics assistance, the secretariat and all those who contributed so efficiently to the
successful organization of this Workshop.

\medskip
      
\begin{center}
                                   
With compliments and kind regards,

\bigskip  
                                            
Hector J de Vega, Norma G Sanchez

\end{center}

\newpage

\section{Programme and Lecturers}

\begin{itemize}

\item{{\bf Peter BIERMANN} (MPI-Bonn, Germany \& Univ of Alabama, Tuscaloosa, USA) 
Astrophysical Dark Matter, keV scale particle and fermion condensates.}

\item{{\bf Alfonso CAVALIERE} (Dipt Fisica/Astrofisica, Unuv Roma 2 Tor Vergata, Italy) 
The Intra Cluster Medium in Dark Matter Halos }

\item{{\bf Hector J. DE VEGA} (CNRS LPTHE Univ de Paris VI, France) 
Galaxy properties from linear primordial fluctuations and 
keV scale dark matter from theory and observations} 

\item{{\bf Gianfranco GENTILE} (Math Physics \& Astronomy, Univ of Ghent, Ghent, Belgium) S
urface densities and dark matter properties in galaxies} 

\item{{\bf Yehuda HOFFMAN} (Racah Inst of Physics, Hebrew Univ, Jerusalem, Israel) 
Dark matter halos with and without baryons }

\item{{\bf Chanda J. JOG}, Department of Physics, Indian Institute of Science, 
Bangalore, IndiaDetermination of the density profile of the 
dark matter halos in galaxies}

\item{{\bf Anatoly KLYPIN} (Dept. of Astronomy, New Mexico State University, USA ) 
Stefan GOTTLOBER(Astr Inst Postdam, Postdam, Germany) 
Dark matter halos from N-body simulations. 
Bolshoi N-body Cosmological Simulation }

\item{{\bf Andrea LAPI} (Dipt Fisica/Astrofisica, Univ Roma 2 Tor Vergata, Rome, Italy) 
Probing Dark Matter Halos in Galaxies and their Clusters} 

\item{{\bf Jounghun LEE} (Dep. Phys. \& Astronomy, Seoul National University, 
Seoul, South Korea) 
Bullet Clusters and its Cosmological Implications }

\item{{\bf Andrea V. MACCIO} (Max Planck Institut fur Astronomie, Heidelberg, Germany) 
Dark Matter at small scales: the lesson from Milky Way satellites} 

\item{{\bf Paolo SALUCCI} (SISSA-Astrophysics, Trieste, Italy) 
Universality Properties in Galaxies and Cored density Profiles }

\item{{\bf Norma G. SANCHEZ} (CNRS LERMA Observatoire de Paris, Paris, France) 
Galaxy properties, keV scale dark matter from theory and observations 
and the power of linear approximation }

\item{{\bf Pasquale D. SERPICO} (CERN-Theory Division \& LAPTH Annecy-le-Vieux, France) 
Astrophysical explanations of the cosmic positron excess in 
Pamela and Fermi }

\item{{\bf Rainer STIELE} (Inst Theor Phys, Heidelberg University, Heidelberg, Germany) 
Cosmological bounds on dark matter self-interactions} 

\item{{\bf Janine VAN EYMEREN} (Univ of Manchester UK \& Duisburg-Essen, Germany) 
Non-circular motions and the Cusp/Core discrepancy in dwarf galaxies }

\item{{\bf Matthew G. WALKER} (Institute of Astronomy, University of Cambridge, UK) 
A universal mass profile for dwarf spheroidal galaxies.}

\item{{\bf Markus WEBER} (Inst fur Experimentelle Kemphysik, Karlsruher Inst. fur 
Technologie KIT, Karlsruhe, Germany) 
The determination of the local Dark Matter density }

\item{{\bf Gustavo YEPES} (Grupo de Astrofisica, Univ Autonoma de Madrid, Cantoblanco, Spain) 
How warm can dark matter be ?. Constraining the mass of dark matter 
particles from the Local Universe }

\item{{\bf Gabrijela ZAHARIJAS}, (IphT/CEA-Saclay, Gif-sur-Yvette, France)
Dark matter constraints from the Fermi-LAT observations}

\end{itemize}

\newpage

\section{Highlights by the Lecturers}

\subsection{Peter Biermann$^{1,2,3,4,5}$}

\noindent
{\bf with help from Julia K. Becker$^{6}$, 
Laurentiu Caramete$^{1,7}$, Lou Clavelli$^{3}$, Jens Dryer$^{6}$, Ben 
Harms$^{3}$, Athina Meli$^{8}$, Eun-Suk Seo$^{9}$, \& Todor Stanev$^{10}$}\\

$^{1}$ MPI for Radioastronomy, Bonn, Germany; 
$^{2}$ Dept. of Phys. \& Astron., Univ. of Bonn, Germany ; 
$^{3}$ Dept. of Phys. \& Astr., Univ. of Alabama, Tuscaloosa, AL, USA; 
$^{4}$ Dept. of Phys., Univ. of Alabama at Huntsville, AL, USA; 
$^{5}$ Dept. of Phys., Karlsruher Institut f{\"u}r Technologie KIT, 
Germany, 
$^{6}$ Dept. of Phys., Univ. Bochum, Bochum, Germany; 
$^{7}$ Institute for Space Sciences, Bucharest, Romania; 
$^{8}$ ECAP, Physik. Inst. Univ. Erlangen-N{\"u}rnberg, Germany; 
$^{9}$ Dept. of Physics, Univ. of Maryland, College Park, MD, USA; 
$^{10}$ Bartol Research Inst., Univ. of Delaware, Newark, DE, USA\\

\vskip0.2cm

\begin{center}

{\bf The nature of dark matter} \\ 

\end{center}

\bigskip

Dark matter has been detected since 1933 (Zwicky) and basically behaves 
like a non-EM-interacting gravitational gas of particles.  From particle 
physics Supersymmetry suggests with an elegant argument that there 
should be a lightest supersymmetric particle, which is a dark matter 
candidate, possibly visible via decay in odd properties of energetic 
particles and photons:  

\medskip

We have discovered i) an upturn in the 
CR-positron fraction, ii) an upturn in the CR-electron spectrum, iii) a 
flat radio emission component near the Galactic Center (WMAP haze), iv) 
a corresponding IC component in gamma rays (Fermi haze), v) the 511 keV 
annihilation line also near the Galactic Center, and most recently, vi) 
an upturn in the CR-spectra of all elements from Helium.  

\medskip

All these 
features can be quantitatively explained with the action of cosmic rays 
accelerated in the magnetic winds of very massive stars, when they 
explode (Biermann et al. 2009, 2010), based on predictions from 1993 
(Biermann 1993, Biermann \& Cassinelli 1993, Biermann \& Strom 1993, 
Stanev et al 1993).  This allows to go back to galaxy data to derive the 
key properties of the dark matter particle: Work by Hogan \& Dalcanton 
(2000), Gilmore et al. (from 2006, 2009), Strigari et al. (2008), and 
Boyanovsky et al. (2008) clearly points to a keV Fermion particle.  

\medskip

A right-handed  sterile neutrino is a candidate to be this particle 
(e.g. Kusenko \& Segre 1997; Fuller et al. 2003; Kusenko 2004; for a review see 
Kusenko 2009; Biermann \& Kusenko 2006; Stasielak et al. 2007; 
Loewenstein et al. 2009): 

\medskip

This particle has the advantage to allow star 
formation very early, near redshift 80, and so also allows the formation 
of supermassive black holes, possibly formed out of agglomerating 
massive stars.  

\medskip

Black holes in turn also merge, but in this manner start 
their mergers at masses of a few million solar masses.  This readily 
explains the supermassive black hole mass function.  The corresponding 
gravitational waves are not constrained by any existing limit, and could 
have given a substantial energy contribution at high redshift.  

\medskip

Our conclusion is that a right-handed  sterile 
neutrino of a mass of a few keV is the 
most interesting candidate to constitute dark matter.

\bigskip

{\bf Acknowledgements:}

\medskip

PLB would like to thank G. Bisnovatyi-Kogan, J. Bl{\"u}mer, R. Engel, 
T.K. Gaisser, L. Gergely, G. Gilmore, A. Heger, G.P. Isar, P. Joshi, 
K.H. Kampert, Gopal-Krishna, A. Kusenko, N. Langer, M. Loewenstein, I.C. 
Mari\c{s}, S. Moiseenko, B. Nath, G. Pavalas, E. Salpeter, N. Sanchez, 
R. Sina, J. Stasielak, V. de Souza, H. de Vega, P. Wiita, and many 
others for discussion of these topics.  Support for ESS comes from NASA 
grant NNX09AC14G and for TS comes from DOE grant UD-FG02-91ER40626.

\newpage

\subsection{Alfonso Cavaliere \& Andrea Lapi}

\begin{center}
Dip. Fisica, Univ. `Tor Vergata', Via Ricerca
Scientifica 1, 00133 Roma, Italy,

SISSA, Via Bonomea 265, 34136 Trieste, Italy

\medskip

and

\medskip

{\bf R. Fusco-Femiano}, 
INAF-IASF, Via Fosso del Cavaliere, 00133 Roma, Italy.

\bigskip

{\bf The Intra Cluster Medium in Dark Matter Halos}

\end{center}

\medskip

In galaxy clusters the gravitational potential wells set by
dark matter (DM) masses $ M\sim 10^{15}\,M_{\odot} $ are filled
out to the virial radius $ R\sim $ Mpc by a hot thin medium at
temperatures $k_BT\sim $ several keVs, with central particle
densities $n\sim 10^{-3}$ cm$^{-3}$.

Such a medium constitutes a remarkably good electron-proton
plasma (appropriately named IntraCluster Plasma, ICP), with a
huge ratio of thermal to mean electrostatic energy $k_B T/e^2\,
n^{1/3}\sim 10^{12}$. It emits copious X-ray powers $L_X\propto
n^2\, T^{1/2}\, R^3\sim 10^{45}$ erg s$^{-1}$ via thermal
bremsstrahlung; but over most of the cluster volume the large
thermal energy content $ E\propto n\,k_BT\, R^3\approx
10^{63-64} $ ergs makes the radiative cooling time longer than
the cluster age.

From the macroscopic viewpoint, the ICP constitutes a simple
fluid with $3$ degrees of freedom and effective particle mass
$ \mu m_p\approx 0.6\,m_p $ in terms of the proton's $ m_p $. So it
grants precision modeling as for the space distributions of
density $n(r)$ and temperature $ T(r) $, such as to match the
rich amount of current data concerning the emissions in X rays,
and the upcoming measurements of the Sunyaev-Zel'dovich
scattering in $\mu$waves.

Handy and effective modeling is provided by the Supermodel.
This is based on the run of the ICP `entropy' (adiabat)
$k\equiv k_B T/n^{2/3}$ provided the physical processes for its
production. The entropy is raised both at the cluster centers
due to the energy discharged by deep mergers and AGN outbursts,
and at the virial boundary from shocking the gravitational
inflow of external gas accreted along with the DM. These
processes together originate ICP entropy profiles with shapes
$k(r)=k_c+(k_R-k_c)\,(r/R)^a$ comprising a central floor
$k_c\approx 10-100$ keV cm$^2$, and an outer ramp with slope
$a\approx 1.1$ adjoining to the boundary values $k_R\sim$ a few
$10^3$ keV cm$^2$, consistent with the recent analyses of wide
cluster samples in X rays.

The ensuing gradient of the thermal pressure $ p(r)\propto
k(r)\,n^{5/3}(r)$  is used in the Supermodel to balance the DM
gravitational pull $ -G\,M(<r)/r^2 $, and sustain hydrostatic
equilibrium. The latter is solved to \emph{directly} yield the
temperature profile in terms of the entropy run $k(r)$. Density
and temperature are \emph{linked} together by $ n(r)=[k_B
T(r)/k(r)]^{3/2} $, so the related X-ray observables are readily
derived and compared with data. With the three specific
parameters appearing in $k(r)$ the Supermodel has provided
remarkably good fits to the X-ray data on surface brightness
and temperature profiles for several galaxy clusters. The fits
not only include the central morphologies from cool-core to
non-cool-core (CCs and NCCs), but also cover diverse outer
behaviors including the steep temperature profiles recently
observed. The Supermodel also enables us to derive from X rays
the the DM concentration $c$; this yields the cluster Grand
Design illustrated in Fig.~1. The interested reader may himself
try more clusters on using the fast Supermodel algorithm made
available at the website
\textsl{http://people.sissa.it/$\sim$lapi/Supermodel/}.

Currently we are working toward including the additional
support to ICP equilibrium provided by turbulence; this is
driven by inflows of intergalactic gas across the cluster
boundary and past the accretion shocks. We find (consistently
with X-ray observations) that such phenomena are increasingly
important at low redshifts, when the shocks weaken and the
infall itself subside due to the accelerating Universe and the
feeding on the initial perturbation wings.

Thus the Supermodel is proving to be an interesting and handy
tool to model and probe turbulence amplitude and decay scale in
a plasma. This constitutes an enticing if complex astrophysical
issue, such to warrant close \emph{modeling} and
\emph{probing}.

\begin{description} 
\item{1} Cavaliere, A., Lapi, A., and Fusco-Femiano, R.
2009, ApJ, 698, 580.
\item{2} Cavaliere, A., Lapi, A., and Fusco-Femiano, R.
2009, ApJ, 698, 580.
\item{3} Fusco-Femiano, R., Cavaliere, A., and Lapi, A.
2009, ApJ, 705, 1019.
\item{4} Lapi, A., Cavaliere, A., and Fusco-Femiano, R.
2010, A\&A, 516, A34.
\end{description}

\begin{figure}[t]
\includegraphics[scale=0.8]{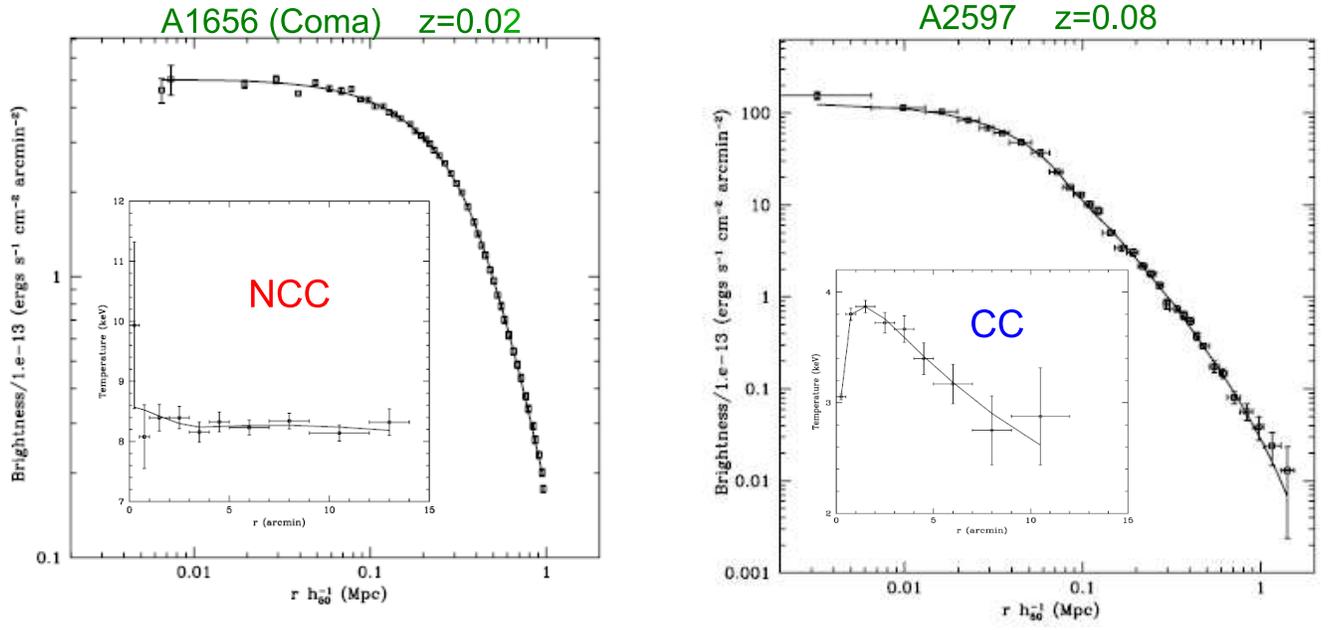}
\caption{The cluster Grand
Design resulting from development of the DM halos and
ICP analyses with the Supermodel, is illustrated with examples
of X-ray brightness profiles (projected temperatures in the
insets). Outer production modulates the outskirts, while
central entropy marks the CC/NCC dichotomy, which we find
correlated to high/low DM concentrations.}
\end{figure}

\newpage

{} {} $ \; \; {} $

\newpage
\subsection{Hector J. de Vega}

\vskip -0.3cm

\begin{center}

LPTHE, CNRS/Universit\'e Paris VI-P. \& M. Curie \& Observatoire de Paris.

\bigskip

{\bf Galaxy properties from linear primordial fluctuations and 
keV scale dark matter from theory and observations} 

\end{center}

The Standard $\Lambda$CDM Cosmological Model begins by the {\bf inflationary} 
era, slow-roll inflation explains the horizon and flatness features of the
present Universe and gravity is described by Einstein's General Relativity.
Particle Physics is described by the Standard Model of particle physics:
the $ SU(3) \otimes SU(2) \otimes U(1) = $ (qcd+electroweak model).
Dark matter must be {\bf cold} (non-relativistic) when
structure formation happens. DM is outside the SM of particle physics.
Finally, the dark energy is described by the cosmological constant $ \Lambda $.
The standard Cosmological model has been validated by a huge host of data
of completely different nature obtained by independent methods
from many kinds of astrophysical and cosmological observations.

In the context of the standard Cosmological model the nature of DM
is unknown. However, it is a forefront problem of modern cosmology
since $ 83\% $ of the matter in the universe is dark.
Only the DM gravitational effects are noticed and they are necessary
to explain the present structure of the Universe.

DM (dark matter) particles must be neutral and so weakly interacting 
that no effects are so far detectable.
Extremely many candidates in particle physics models beyond the standard model
of particle physics.

Theoretical analysis combined with astrophysical data from
galaxy observations points towards a DM particle mass in
the {\bf keV scale} (keV = 1/511 electron mass) [1-4].

DM particles can decouple being ultrarelativistic (UR) at 
$ \; T_d \gg m $ or non-relativistic $ \; T_d \ll m $.
They may  decouple at or out of local thermal equilibrium (LTE).

The DM distribution function: $ F_d[p_c] $ freezes out at decoupling
becoming a function of the  comoving momentum $ p_c = $.
$ P_f(t) = p_c/a(t) = $ is the physical momentum. 

Basic physical quantities can be expressed in terms of the
distribution function as the velocity fluctuations,
\be\label{uno}
\langle \vec{V}^2(t) \rangle = \langle \frac{\vec{P}^2_f(t)}{m^2} \rangle = 
 \left[\frac{T_d}{m \; a(t)}\right]^2  \; 
\frac{\int_0^\infty y^4 F_d(y) dy}{\int_0^\infty y^2 F_d(y) dy} 
\ee
and the DM energy density,
\be\label{dos}
\rho_{DM}(t) = \frac{m~g}{2\pi^2} \; \frac{T^3_d}{a^3(t)}
\int_0^{\infty} y^2  \;  F_d(y)  \; dy \; ,
\ee
where $ y = P_f(t)/T_d(t) = p_c/T_d $ is the integration variable and
$ g $  is the number of internal degrees of freedom of the DM 
particle; typically $ 1 \leq g \leq 4 $.

{\bf Two} basic quantities characterize DM: its particle mass $ m $
and the temperature $ T_d $ at which DM decouples. This last quantity
is related by entropy conservation to the number of
ultrarelativistic degrees of freedom $ g_d $ at decoupling by
$ \quad T_d = \left(\frac2{g_d}\right)^\frac13 \; T_{cmb} \; ,
\; T_{cmb} = 0.2348 \; 10^{-3} \; $ eV.
Notice that $F_d(y)$ is of order one and that eqs.(\ref{uno}) and (\ref{dos})
are valid all the time before structure formation.

\medskip

One therefore needs {\bf two} constraints to determine the values of
$ m $ and $ T_d $ (or $ g_d $).

\medskip

One constraint is to reproduce the known cosmological DM density today.
$\rho_{DM}({\rm today})= 1.107 \; \frac{\rm keV}{{\rm cm}^3}$.

Two independent further constraints are considered in refs. [1-4].
First, the phase-space density $ Q=\rho/\sigma^3 $ [1-2] and second the
surface acceleration of gravity in DM dominated galaxies [3-4].
We therefore provide {\bf two} quantitative ways to derive the value $ m $ 
and $ g_d $ in refs. [1-4].

\medskip

The phase-space density $ Q $ is invariant under the
cosmological expansion and can {\bf only decrease} 
under self-gravity interactions 
(gravitational clustering). The value of $ Q $ today follows
observing dwarf spheroidal satellite galaxies of the Milky Way (dSphs):
$ Q_{today} = (0.18 \;  \mathrm{keV})^4 $ (Gilmore et al. 07 and 08).
We compute explicitly $ Q_{prim} $ (in the primordial universe) and it turns
to be proportional to $ m^4 $ [1-4].

During structure formation $ ( z \lesssim 30 ), \;  Q $
{\bf decreases} by a factor that we call $ Z $. Namely, 
$ Q_{today} = Q_{prim}/Z $. The value of $ Z $ is galaxy-dependent.
The spherical model gives $ Z \simeq 41000 $
and $N$-body simulations indicate: $ 10000 >  Z > 1 $ (see [1]).

Combining the value of $ Q_{today} $ and $\rho_{DM}({\rm today}) $ with 
the theoretical analysis yields that $ m $ must be in the keV scale and 
$ T_d $ can be larger than 100 GeV. More explicitly, we get from 
eqs.(\ref{uno}) and (\ref{dos}) general formulas for $ m $ and $ g_d $:
$$ 
m = \frac{2^\frac14 \; \sqrt{\pi}}{3^\frac38 \; g^\frac14 } \; 
Q_{prim}^\frac14
\; I_4^{\frac38} \; I_2^{-\frac58} \; , \quad
g_d = \frac{2^\frac14 \; g^\frac34}{3^\frac38 \; 
\pi^\frac32 \; \Omega_{DM}} \; 
 \; \frac{T_{\gamma}^3}{\rho_c} \; Q_{prim}^\frac14 \; 
\left[I_2 \; I_4\right]^{\frac38}
$$
where $ I_{2 \, n} = \int_0^\infty y^{2 \, n} \; F_d(y) \; dy 
\quad , \quad n=1, 2 $ 
and $ Q_{prim}^\frac14 = Z^\frac14 \; \; 0.18 $ keV using the dSphs data,
$T_{\gamma} = 0.2348 \; {\rm meV } 
\; , \; \Omega_{DM} = 0.228 $ and $ \rho_c = (2.518 \; {\rm meV})^4$.

These formulas yield for relics decoupling UR at LTE:
$$ 
m = \left(\frac{Z}{g}\right)^\frac14 \; \mathrm{keV} \; 
\left\{\begin{array}{l}
         0.568 \\
              0.484      \end{array} \right. \; , \;
 g_d = g^\frac34 \; Z^\frac14 \; \left\{\begin{array}{l}
         155~~~\mathrm{Fermions} \\
              180~~~\mathrm{Bosons}      \end{array} \right. \; . 
$$
Since $ g = 1-4 $, we see that 
$ g_d \gtrsim 100 \Rightarrow  T_d \gtrsim 100 $ GeV.
Moreover, $ 1 < Z^\frac14 < 10 $ for $ 1 < Z < 10000 $.
For example for DM Majorana fermions $ (g=2) \; m \simeq 0.85 $ keV.

Results for $ m $ and $ g_d $ on the same scales for DM particles 
decoupling UR out of thermal equilibrium [1].

For a specific model of sterile neutrinos where decoupling is out of 
thermal equilibrium:
$$
0.56 \; \mathrm{keV} \lesssim m_{\nu} \;  
Z^{-\frac14} \lesssim 1.0 \; \mathrm{keV}
\quad ,  \quad 15 \lesssim g_d  \;  Z^{-\frac14}\lesssim 84
$$
So far we considered UR decoupling. For relics decoupling non-relativistic
we obtain similar results for the DM particle mass: keV 
$ \lesssim m \lesssim $ MeV [1].

\medskip

\begin{figure}
\begin{turn}{-90}
\psfrag{"pwimp.dat"}{}
\psfrag{"p10ev.dat"}{}
\psfrag{"p1kev.dat"}{}
\includegraphics[height=9.cm,width=6.cm]{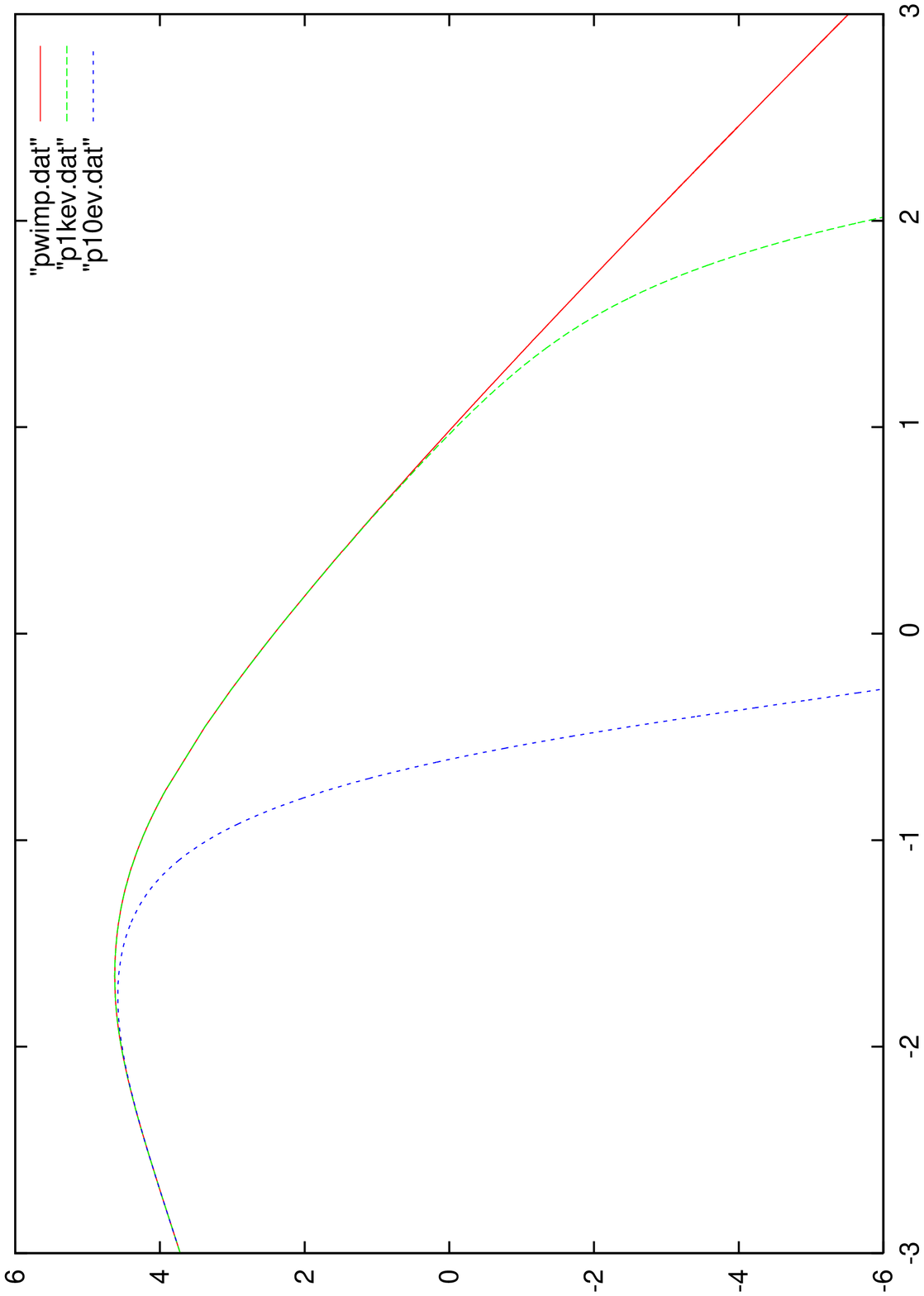}
\end{turn}
\caption{}
\label{poten}
\end{figure}

The value of the DM particle mass affects the linear primordial power today 
for small scales. We plot in fig. \ref{poten} $ \log_{10} P(k) $ vs. 
$ \log_{10}[k \;  {\rm Mpc}  \; h] $ for WIMPS (red),  
1 keV DM particles (green)
and 10 eV DM particles (blue). Recall that 
$ P(k) = P_0 \; k^{n_s} \; T^2(k) $
where $ T(k) $ is the transfer function in the MD era that
we computed  from the Gilbert integral equation [5].
The power $ P(k) $ turns to be cutted for 1 keV DM particles on scales 
$ \lesssim 100 $ kpc. For scales {\bf larger} than $ \sim 100 $
kpc, DM particles and wimps give {\bf identical} results.
10 eV DM particles are ruled out because they  suppress all structures
below $ \sim 3 $ Mpc.

\medskip

Many extensions of the SM of particle physics 
include a DM particle with mass in the {\bf keV scale} and
weakly enough coupled to the SM particles to fulfill
all particle physics experimental constraints. 
Main candidates in the keV mass scale: sterile neutrinos, 
light gravitinos, light neutralino, majoron ...

The proposal of sterile neutrinos is motivated by the fact
that there are both left and right handed quarks.
It is then natural to have right handed neutrinos $ \nu_R $ 
besides the known left-handed neutrino (quark-lepton similarity).

Sterile neutrinos can transmute into ordinary neutrinos
and viceversa through mixing (non-diagonal mass terms). 
For $ m_{{\rm sterile} \nu} \sim 1 $ keV and 
$ m_{{\rm ordinary} \; \nu} \sim 0.1 $ eV the 
mixing angle turns to be $ \theta \sim 10^{-4} $. This small value
is appropriate to produce enough sterile neutrinos to
account for the observed DM.
Smallness of $ \theta $ makes very difficult to detect steriles.  
The most promising experiments are those of beta decay
where a sterile should be produced instead of an ordinary
neutrino with probability $ \sim \theta^2 $. In particular,
renium and tritium beta decay are the best candidates since they
provide the lowest energy yield.

\medskip

{\bf References}

\begin{description}
\item[1]  H. J. de Vega, N. G. Sanchez,  arXiv:0901.0922, 
Mon. Not. R. Astron. Soc. 404, 885 (2010).
\vspace{-0.2cm}
\item[2] D. Boyanovsky, H. J. de Vega, N. G. Sanchez, 	
arXiv:0710.5180, Phys. Rev. {\bf D 77}, 043518 (2008).
\vspace{-0.2cm}
\item[3] H. J. de Vega, N. G. Sanchez, arXiv:0907.0006.
\vspace{-0.2cm}
\item[4] H. J. de Vega, P. Salucci, N. G. Sanchez, arXiv:1004.1908.
\vspace{-0.2cm}
\item[5] D. Boyanovsky, H. J. de Vega, N. G. Sanchez, 	
arXiv:0807.0622, Phys. Rev. {\bf D 78}, 063546 (2008).
\end{description}

\newpage

\subsection{Gianfranco Gentile}

\vskip -0.2cm

\begin{center} 
University of Ghent, Belgium.

\bigskip

\bigskip

{\bf Surface densities and dark matter properties in galaxies}

\end{center}

\vspace{0.6cm}
 
Rotation curves of spiral galaxies 
(their rotation velocity as a function of galactocentric radius) 
do not decline as expected from the observed distribution of matter. 
Two proprosed solutions are either to 
envisage that galaxies are embedded in a halo of yet-to-be-discovered 
particles (a dark matter halo), 
or to consider the possibility (e.g. MOND, Modified Newtonian Dynamics, Milgrom 1983) that 
gravity does not behave exactly as we would expect from the extrapolation
to very weak fields of the known Newton (or Einstein) gravities.

The current standard framework of formation of structures in the Universe is the so-called 
$\Lambda$ Cold Dark Matter ($\Lambda$CDM) framework. 
Standard (dark matter only) $\Lambda$CDM 
simulations of
structure formation in the Universe result in dark matter halos characterised by a central
density ``cusp'' ($\rho \propto r^{-1}$ for small radii; the exact value of the asymptotic
inner slope is still a matter of debate), 
whereas the observations (e.g. Gentile et al. 2005, 2007; van Eymeren et al. 2009;
de Blok 2010, and references therein) tend to favour central constant density cores.
Systematic effects that would invalid this conclusion (e.g., non-circular motions,
resolution effects) seem to be under control, thanks to improved observational
and analysis techniques.
From the theoretical point of view, the effect of baryons on dark matter structures 
is not trivial to implement. The best
understood effect of baryons is adiabatic contraction (e.g., Blumenthal
et al. 1986, Sellwood \& McGaugh 2005), where the dark matter halo however becomes even more 
centrally concentrated (and therefore even more in contrast with observations) 
as a result of baryons cooling and infall in the central parts
of the dark matter halo. On the other hand, some groups (e.g.
Mashchenko et al. 2008, Governato et al. 2010) have produced simulations where 
approximately constant density cores are formed, as a result of feedback processes linked
to star formation. However, consensus is far from being reached on how to implement
baryonic physics in dark matter simulations.

One of the most widely used functional forms for a cored halo is the so-called Burkert halo
(Burkert 1995):
\begin{equation}   
\rho_{\rm Bur}(r)=\frac{\rho_0 r_{0}^3}{(r+r_{0})   
(r^2+r_{0}^2)},   
\end{equation}      
where $\rho_0$ is the central density and $r_{0}$ is the core radius.    

Donato et al. (2009) analysed a sample of galaxies (with published mass models) of 
all Hubble types and spanning 14 galaxy magnitudes. They reached the conclusion (already
noted by Kormendy \& Freeman 2004 and Spano et al. 2008 
for smaller samples and with fewer galaxy
types) that
the product $\rho_0 r_{0}$ is approximately the same for all galaxies: 
$\rho_0 r_{0}$=141$^{+82}_{-52}$ M$_{\odot}$ pc$^{-2}$. This is equivalent to 
the average dark matter 
surface density within $r_0$: 
$<\Sigma>_{0,DM}=M(< r_0)/(\pi r_0^2) 
\sim 0.51 \rho_0 r_0 = 72^{+42}_{-27}$~M$_\odot$~pc$^{-2}$, 
which is also equivalent to the gravitational acceleration generated by
dark matter at $r_0$: $g_{DM}(r_0) = G \pi <\Sigma>_{0,DM} 
= 3.2^{+1.8}_{-1.2} \, 10^{-9}$cm~s$^{-2}$.

In Gentile et al. (2009) we found that also the {\it baryonic} surface density within $r_0$
is universal, see Fig. 1. When expressed in terms of the gravitational 
acceleration generated by baryons
at $r_0$, this universality reads: $g_{b}(r_0) = 5.7^{+3.8}_{-2.8} \, 10^{-10}$ cm s$^{-2}$.

We note that these universal relations hold at $r_0$, even though within the galaxies of
our sample the {\it central} surface density of baryons varies by more than four orders
of magnitude (Donato et al. 2009).

The interpretation of these universal relations is far from being simple. 
However, these relations seem to point to the idea that dark matter ``knows'' what 
baryons are doing: at $r_0$, the gravitational acceleration
due to dark matter and the gravitational acceleration due to baryons are 
about the same for every galaxy. From the gravitational field of baryons,
one can derive (roughly) dark matter properties.

If the dark matter particle has a mass around 1-2 keV, then de Vega, 
Salucci \& Sanchez (2010)
have shown that it can reproduce the observed universality of dark matter surface density.

\begin{figure}[tbh]
\includegraphics[scale=0.7]{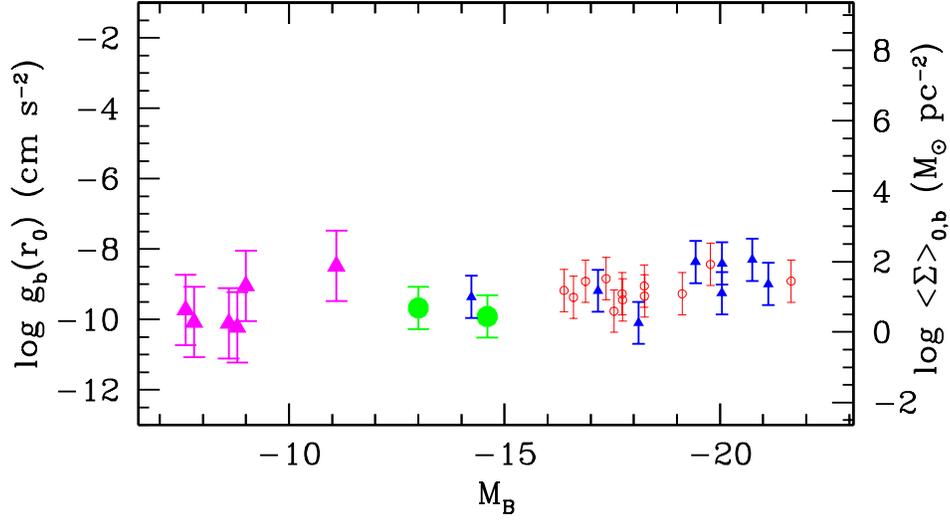}
\caption{
Universality of the average surface density (and gravity) 
of baryons within the halo core radius
(plotted against B-band absolute magnitude; from Gentile et al. 2009). Different symbols
indicate different galaxy subsamples. 
}
\end{figure}

\bigskip

{\bf References}

\medskip

{\small \noindent
Blumenthal, G.~R., Faber, S.~M., Flores, R., \& Primack, J.~R.\ 1986, ApJ, 301, 27 \\
Burkert, A.\ 1995, ApJ, 447, L25 \\
de Blok, W.~J.~G.\ 2010, Advances in Astronomy, 2010 \\
de Vega, H.~J., Salucci, P., \& Sanchez, N.~G.\ 2010, arXiv:1004.1908 \\
Donato, F., et al.\ 2009, MNRAS, 397, 1169 \\
Gentile, G., Burkert, A., Salucci, P., Klein, U., \& Walter, F.\ 2005, ApJ, 634, L145 \\
Gentile, G., Salucci, P., Klein, U., \& Granato, G.~L.\ 2007, MNRAS, 375, 199 \\
Gentile, G., Famaey, B., Zhao, H., \& Salucci, P.\ 2009, Nature, 461, 627 \\
Governato, F., et al.\ 2010, Nature, 463, 203 \\
Kormendy, J., \& Freeman, K.~C.\ 2004, Dark Matter in Galaxies, 220, 377 \\
Mashchenko, S., Couchman, H.~M.~P., \& Wadsley, J.\ 2006, Nature, 442, 539 \\
Milgrom, M.\ 1983, ApJ, 270, 365 \\
Sellwood, J.~A., \& McGaugh, S.~S.\ 2005, ApJ, 634, 70 \\
Spano, M., Marcelin, M., Amram, P., Carignan, C., Epinat, B., 
\& Hernandez, O.\ 2008, MNRAS, 383, 297 \\
van Eymeren, J., Trachternach, C., Koribalski, B.~S., \& 
Dettmar, R.-J.\ 2009, A\&A, 505, 1 }

\newpage

\subsection{Chanda J. Jog}

\centerline {Indian Institute of Science, Bangalore 560012, India}

\bigskip

\bigskip

\centerline {\bf Tracing the dark matter halos of galaxies by modeling
                the observed HI data}

\bigskip

\centerline{\bf Abstract}

\medskip

We use the observed rotation curves and the HI vertical scaleheight
data as simultaneous constraints to obtain the density profile and 
the shape of the dark matter halos in galaxies, and show these to be 
varied. A galaxy is modeled 
as a gravitationally coupled star-gas system in the field of the 
dark matter halo. This approach is applied to three galaxies:
the Milky Way, M31 and UGC 7321, a low surface brightness
galaxy. The resulting dark matter profile is not 
universal.

\bigskip

\centerline {\bf Introduction}

\medskip

In a typical spiral galaxy like our Galaxy, stars constitute $\sim 90 \%$
of the visible mass while the rest is in the interstellar gas. Due to its 
lower dispersion, gas is important for the disk dynamics despite its lower mass 
content. Further, the interstellar atomic hydrogen gas (HI) extends 2-3  times 
farther out than the stellar disk, hence it  is an excellent  tracer of 
dynamics in the outer regions of a galaxy.

The observed rotation curve has been routinely used in the literature to 
study the radial mass distribution within a galaxy, and it is well-known 
that the dark matter progressively dominates in the outer parts. However,
the shape of the dark matter halo is not well-studied. 
We use the observed vertical 
thickness of the HI gas distribution as an additional, 
complementary constraint  
to model the vertical density distribution and hence the shape of the halo.

\bigskip

\centerline {\bf Calculations and Results}

\medskip

A galactic disk is supported vertically by pressure, and the 
balance of self-gravity and  pressure decides its  thickness. This was studied
for a one-component, gravitating isothermal disk 
in a classic paper by Spitzer (1942).
However, a real galaxy consists of stars and gas, where the 
gas gravity could play a crucial role in 
determining the vertical distribution since the gas lies 
closer to the mid-plane. 
To study this, we have developed a model where stars and gas are 
taken to be gravitationally coupled, and are embedded in the field 
of a rigid dark matter halo  (Narayan \& Jog 2002).

We solve the equations of hydrostatic equilibrium along $z$  for stars and gas 
and the joint Poisson equation  together to obtain a self-consistent solution 
for the vertical disk density distribution.   The coupled, second-order differential 
equations are solved   numerically and iteratively. A general four-parameter 
dark-matter halo profile as motivated from the dynamical studies of elliptical 
galaxies (de Zeeuw \& Pfenniger 1988) is used where the four parameters 
are: $\rho_0$, the central density; $R_c$, the core radius;  $q$, the vertical to 
planar axis ratio; and   $p$, the power-law density index. The grid of halo 
parameters is scanned systematically. For different trial halo parameters, we 
obtain the rotation curve and the vertical scaleheights at different radii 
using our model. We then compare these with the observations to get the best-fit  
halo parameters. This approach has been applied to study three galaxies and the 
results are summarized in Table 1.

For the nearby Andromeda galaxy, our best-fit 
model gives an isothermal, flattened halo with an axis ratio of 
0.4 (Banerjee \& Jog 2008). This lies at the most oblate end of 
the distribution obtained from cosmological simulations. 
For the low surface brightness, superthin 
galaxy UGC 7321, we find that the best-fit halo-core radius is comparable 
to the stellar disk scalelength. Thus the dark matter halo dominates 
the dynamics even at small radii in a LSB galaxy  (Banerjee, Matthews \&
 Jog 2010). For the Galaxy, the best-fit to the HI data 
indicates a spherical halo with a density falling faster than for an 
isothermal case (Narayan, Saha \& Jog 2005).
Thus, the dark matter halo profile in spiral galaxies does not appear 
to be universal.

The three crucial parameters that determine the disk vertical HI distribution  are the
disk-to-halo mass ratio, the gas velocity dispersion, and the shape of the halo.
Our work underlines the potential of this approach for
 further systematic study of the dark matter halo parameters
 in different galaxy types. For this we need data for outer-galactic HI scaleheights 
 \& gas dispersion - this remains a challenging task
  as was already pointed out by Sancisi \& Allen (1979)
 in their study of NGC 891.

The shape of the dark matter halo in the disk plane is also 
 an important parameter in the study of galaxy formation and evolution. 
However, this has not been studied much so far. 
Recent, detailed Leiden/Argentine/Bonn HI survey of the 
Milky Way Galaxy shows the gas thickness to 
be azimuthally asymmetric with the flaring being much higher in the North. 
We assume the gas to be in a hydrostatic equilibrium, and model the above 
asymmetry of thickness
to constrain the shape 
of the halo in the disk plane. We conclude that the Galactic dark
matter halo is lopsided and elongated in the disk plane (Saha et al. 2009). 

\bigskip

\noindent{\bf References:}  

\medskip

\noindent Banerjee, A.,  \& Jog, C.J. 2008, ApJ, 685, 254

\noindent Banerjee, A., Matthews, L.D.,  \& Jog, C.J. 2010, New Astron., 15, 89

\noindent de Zeeuw, T., \& Pfenniger, D. 1988, MNRAS, 235, 949

\noindent Narayan, C.A, \& Jog, C.J. 2002, A \& A, 394, 89

\noindent Narayan, C.J., Saha, K.,  \& Jog, C.J. 2005, A \& A, 440, 523

\noindent Saha, K., Levine, E.S., Jog, C.J., \& Blitz, L. 2009, ApJ, 697, 2015

\noindent Sancisi, R., \& Allen, R.J. 1979, A \& A, 74, 73

\noindent Spitzer, L. 1942, ApJ, 95, 329

\begin{table}
\centering
  \begin{minipage}{140mm}
   \caption{\bf Density profile and shape of best-fit dark matter halo
 }
\begin{tabular}{lll}
\\
Name of galaxy & $\:$ Radial profile & shape, q = vertical to planar axis ratio \\
\\
M 31 & $\:$ 1/R$^2$ & oblate, q=0.4 \\
UGC 7321 & $\:$ 1/R$^{2}$ & spherical, q=1.0 \\
The Galaxy & $\:$ 1/R$^4$ & spherical, q=1.0 \\
\end{tabular}
\end{minipage}
\end{table}

\newpage

\subsection{Paolo Salucci}

\centerline{ SISSA, Trieste, Italy.}

\bigskip

\begin{center}

{\bf Universal Properties in Galaxies and Cored DM Profiles.}

\end{center}
 
\medskip

The presence of large amounts of unseen matter in  galaxies, distributed  
differently from   stars and gas,  is well established   from rotation curves 
which do not show the expected Keplerian fall-off at large radii,  but remain 
increasing, flat or start to  gently decrease over their observed range. The 
invisible mass component becomes progressively more abundant at outer radii  
and  for the less luminous galaxies (Persic and Salucci and Stel  1996).

In Spirals we have the best opportunity to study the global mass distribution:  the   
gravitational potentials of   a spherical stellar bulge, a dark  halo, a stellar disk 
and  a gaseous disc 
give rise to an observed  equilibrium circular velocity  
 $V^2(r)=r\frac{d}{dr}\phi_{tot}=V^2_b + V^2_{DM}+V^2_*+V^2_{HI}$. 
The Poisson equation  
relates the surface (spatial)  densities of these components to the corresponding  
gravitational potentials. 
The investigation is not difficult: e.g.  $\Sigma_*(r)$,  the surface stellar  
density is proportional (by the  mass-to-light ratio) to the  observed  
surface brightness: $\Sigma_{*}(r)=\frac{M_{D}}{2 \pi R_{D}^{2}}\: e^{-r/R_{D}}$ 
and then $V_{*}^{2}(r)=\frac{G M_{D}}{2R_{D}} x^{2}B\left(\frac{x}{2}\right)$,
 where $M_D$ is the disk mass and $R_D$ is the disk scale length.  

First,  there exists,  at any  galactocentric radii measured in terms of disk 
length-scale $R_n \equiv (n/5)R_{opt}$,  a    {\it radial}  Tully-Fisher relation 
(Yegorova and Salucci 2007)  between  the local rotation  velocity  
$ V_n \equiv V_{rot}(R_n) $ and  the  total galaxy luminosity  
$M_{band} = a_n \log V_n + b_n$. These relationships have a so very 
low scatter that  imply that  Dark and Luminous matter   are coupled.  

PSS and by Salucci et al 2007 \emph{et al.}   have  evidenced  that these systems   
present Universal features in their kinematics  well correlating  with the  global 
galactic properties. 
This has  led to the construction,  from  3200 individual RCs,  of the ``Universal 
Rotation Curve'' of Spirals   $V_{URC}(r; P)$ (see PSS), i.e.  an empirical  
function of   galactocentric  radius $r$, that,  tuned by a global galaxy 
property (e.g. the luminosity), can well reproduce the RC of any object.  
Additional  kinematical  data  and     virial velocities $V_{vir}\equiv 
(G M_{vir}/R_{vir})^{1/2}$,  obtained by  Shankar \emph{et al.} 2006,  
have determined  the URC  out to the virial radii (Salucci \emph{et al.} 2007).  

$V_{URC}$  is the  observational counterpart of the  velocity profile 
that emerges  out of numerical cosmological simulations. As individual 
RCs,   it   implies a mass model including  a Freeman disk  and a  
DM halo  with  a Burkert profile $ 
\rho (r)=\frac{\rho_0\, r_0^3}{(r+r_0)\,(r^2+r_0^2)}$.  $r_0$ is the 
core radius and $\rho_0$ the central density,  
see Salucci and Burkert 2000 for details.
  
To assume a cored halo profile is obligatory. It is well known that  
$\Lambda CDM$  scenario provides a successful picture of the cosmological  
structure  formation and that   large N-body  numerical simulations 
performed in  this scenario  lead to the commonly used NFW  halo cuspy  
spatial density profile. However,   a careful analysis  of about 100   
high quality,  extended and  free from deviations from axial symmetry  
RCs has now strongly disfavored   the disk + NFW halo mass model, 
in favor of cored profiles, (e.g. Gentile \emph{et al.} 2004, 2005,     
Spano \emph{et al.} 2007, de Blok 2008 and  de Naray \emph{et al.} 2008).
  
The  structural parameters $\rho_0$,  $r_0$,  $M_D$ are obtained for  the  
URC  and for  any individual RC     by  $\chi^2$ fitting. As result, a cored 
DM distribution and   a set of  scaling laws  among  local and   global galaxy 
quantities    emerges. 

These scaling laws indicate (Salucci \emph{et al.} 2007)  that spirals have  an 
Inner Baryon Dominance region where the stellar disk  dominates  the total 
gravitational potential,  while the   DM halo emerges  farther out.  
At any radii, objects  with lower luminosities have a larger  dark-to-stellar 
mass ratio. The baryonic fraction in a spirals  is always much smaller than the 
cosmological value $\Omega_b/\Omega_{matter} \simeq 1/6  $, and it ranges between 
$ 7\times 10^{-3} $   to   $ 5\times 10^{-2} $,  suggesting that processes such as  
SN  explosions  must have   removed  a very large fraction of the original hydrogen.
Smaller spirals are denser, with their  central density spanning 2 order of 
magnitudes over the mass sequence of spirals.
The stellar mass-to-light ratio   (in the B band)   lies  between 0.5 and 4 
and increase with galaxy  luminosity  as $L_B^{0.2}$;  in  agreement  with the 
values   obtained by fitting the spirals SED with spectro-photometric models.  

\begin{figure}[t!]
\centering
\vskip -0.3cm
\includegraphics[width=9.5truecm]{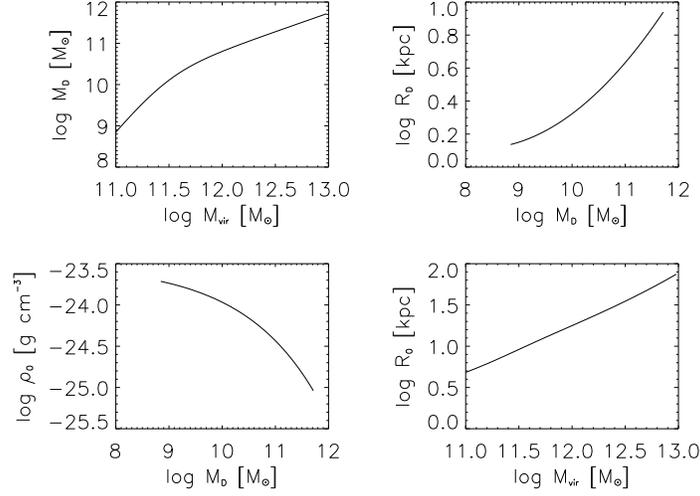}
\vskip -0.6truecm 
\caption{Scaling relations between the structural parameters of the 
dark and luminous mass distribution in spirals.}
\label{fig:scaling_relations}
\vskip -0.5truecm
\end{figure}

As far the structural properties of the DM distribution a  most important   
finding is  that   the central surface density  
$ \propto \mu_{0D}\equiv r_0 \rho_0$,  
where $r_0$ and  $\rho_0$ are  the halo  core radius and central spatial density, is 
nearly constant and  independent of galaxy luminosity.   Based on the co-added 
rotation curves of $\sim 1000$
spiral galaxies, mass models of individual dwarf irregular 
and spiral galaxies of late and early types  with
high-quality rotation curves and, galaxy-galaxy weak lensing signals from a 
sample of spiral and elliptical
galaxies, we find  that   
$ \log \mu_{0D} = 2.15 \pm 0.2$, in units of $\log$(M$_{\odot}$
pc$^{-2}$).  This constancy  transpasses the family of disk systems.  
We also show that the observed  internal kinematics of Local Group dwarf 
spheroidal galaxies,  are
consistent with this value. Our results are obtained for galactic systems 
spanning over 14  magnitudes, 
belonging to different Hubble Types, and whose mass profiles have been 
determined by  several independent 
methods. Very significantly, in  the same objects, the approximate 
constancy of $\mu_{0D}$ is in sharp contrast to  the
systematical  variations, by several orders of magnitude,  of galaxy properties, 
including $\rho_0$  and central  stellar surface density.

The  constancy of $\mu_{0D}$ can be related to the above  
scaling laws of spirals,  as an example,  let us define   $M_{\rm h0}$ and  
$V_{\rm h0}$ is the enclosed 
halo  mass inside $r_0$ and the  halo circular velocity at $r_0$.
 we obtain  $M_{\rm h0} \propto V_{\rm h0}^4$   which immediately reminds 
a sort of Tully-Fisher relation.

The evidence that the DM halo central surface density $\rho_0 r_0$
remains constant to within less than a factor of two over fourteen galaxy magnitudes, and across
several Hubble types,  does indicates  that this quantity  may 
hide  the  physical nature of the DM.Considering that DM haloes are (almost)  
spherical systems it is  surprising that their central surface
density plays  a  role in galaxy structure.  One could wonder whether the physics 
we ``witness''   in the constancy of  $\mu_{0D}$ be instead stored separately in  
$r_0$ and $\rho_0$.  This  interpretation has however a problem:   $r_0$ and 
$\rho_0$ do  correlate with the luminous counterparts, (the disk length-scale 
and stellar central  spatial density) while  $\mu_{0D}$ does not.
Moreover, this evidence, is difficult to  understand   in  an evolutionary 
scenario as  the product of the process that has turned the primordial cosmological  
gas in the stellar structures we  observe today. Such constancy in fact,   
must be achieved in very different  galaxies of different  morphology and mass,   
ranging  from
dark-matter-dominated  to baryon-dominated objects. In
addition, these galaxies have experienced significantly different
evolutionary histories (e.g. numbers of mergers, significance of
baryon cooling, stellar feedback, etc.). 

The best explanation for our findings  relays with the nature itself of the DM, 
as it seems to indicate recent theoretical  work (de Vega et al., 2009,  2010.)  
Then,  the   distribution of matter galaxies  has  become a benchmark for  
understanding dark matter  and the    galaxy formation process. In particular, 
the  universality of certain structural quantities and the dark-luminous coupling  
of the  mass distributions,  seem to  bear the direct imprint of the Nature of 
the  DM (Donato \emph{et al.} 2009, Gentile \emph{et al.} 2009).

\begin{figure}[t!]
\centering
\vskip 0.3truecm
\includegraphics[width=6.2truecm]{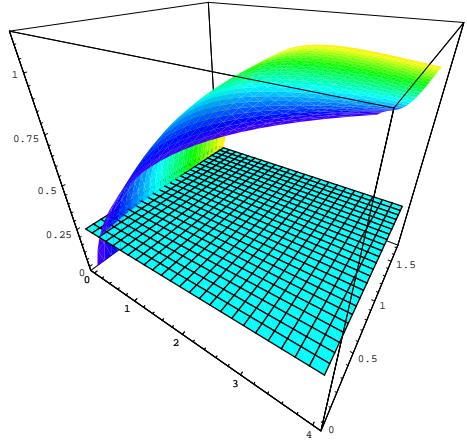}
\vskip -0.1truecm
\caption{The URC. The circular velocity as a function of radius 
(in units of $R_D$) and luminosity (halo mass) see Salucci et al 2007 for details.} 
\label{fig:io_urc1}
\end{figure}

\begin{figure}[t!]
 \vskip -.0truecm
\psfig{file=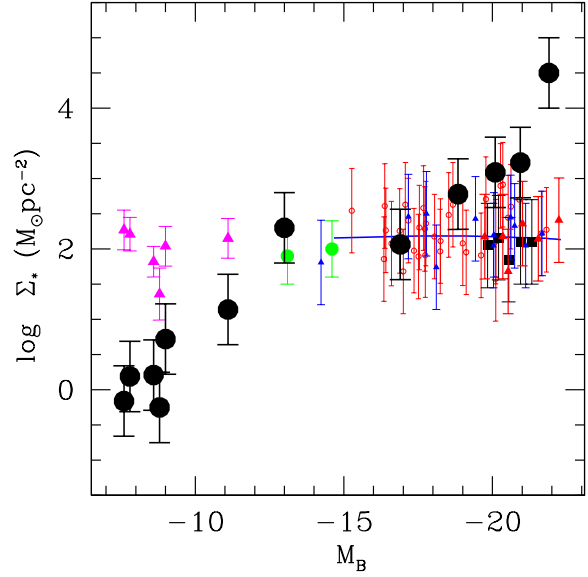,width=0.5\textwidth}   
\vskip -0.8truecm
\caption{Dark matter central surface density in units of 
$M_\odot$pc$^{-2}$  as a function of
galaxy magnitude,  for different galaxies and Hubble Types. 
As  a comparison the values of the same quantity of the stellar 
component is also shown (big filled circles). }
\end{figure}

\begin{description}

\item{1}  
 de Blok,   W.J.G \emph{et al.}  2008, arXiv:0810.2100.

\item{2}
 Donato, F.  Gentile, G. ,  Salucci, P.   2004,
  Mon. Not. Roy. Astron. Soc.,  353, L17.

\item{3} 
Donato F., \emph{et al.} 2009,
Mon. Not. Roy. Astron. Soc., 397, 1169. 
 
\item{4}
 Gentile, G. \emph{et al.}  2004,
 Mon. Not. Roy. Astron. Soc.,  351, 903.
 
\item{5}
Gentile, G.  \emph{et al.} 2005, 
 Astrophys. J.,  634, L145.
 
\item{6}
Gentile G., \emph{et al.} 2009,
Nature, 461, 627. 

\item{7}
 Kuzio de Naray, R. \emph{et al.}   2008, Astrophys. J.,  676, 920.
 
\item{8}
 Navarro,  J.~F., Frenk,   
C.~S., White,  S.~D.~M. 1996, Astrophys. J.,  462, 563 (NFW).
 
\item{9}
Persic, M., Salucci, P. , Stel, F. 1996,
 Mon. Not. R. Astron. Soc.  281, 27 (PSS).
 
\item{10}
Salucci, P. \emph{et al.} 2007,
 Mon. Not. Roy. Astron. Soc.,  378, 41.
 
\item{11}
 Shankar, F. \emph{et al.} 2006, Astrophys. J.,   643, 14.

\item{12}
Spano, M., Marcelin, M., Amram, P. \emph{et al.} 2008, 
Mon. Not. Roy. Astron. Soc.  383, 297.

\item{13}
 Yegorova, I.A.,   Salucci,  P. 2007, Mon. Not. R. Astron. Soc.,  377, 507.

\item{14}
  de Vega, H,  Salucci, P. , Sanchez, N., arXiv:1004.1908 
 
\item{15}
  de Vega, H,   Sanchez N., arXiv0901.0922, Mon. Not. R. Astron. Soc. 404, 885 (2010).
\end{description}

 \newpage
 
\subsection{Norma G Sanchez}

\vspace{-0.3cm}

\begin{center}
Observatoire de Paris, LERMA \& CNRS.

\bigskip

{\bf Galaxy properties, keV scale dark matter from theory and observations 
and the power of linear approximation}
\end{center}

\medskip

Galaxies are described by a variety of physical quantities:

(a) {\bf Non-universal} quantities: mass, size, luminosity, fraction of DM,
DM core radius $ r_0 $, central DM density $ \rho_0 $.

(b) {\bf Universal} quantities: surface density $ \mu_0 \equiv r_0 \; \rho_0 $
and DM density profiles. $M_{BH}/M_{halo}$ (or halo binding energy).

The galaxy variables are related by
{\bf universal} empirical relations. Only one variable remains free. 
That is, the galaxies are a one parameter family of objects.
The existence of such universal quantities may be explained by the
presence of attractors in the dynamical evolution. 
The quantities linked to the attractor always reach the same value
for a large variety of initial conditions. This is analogous
to the universal quantities linked to fixed points in critical
phenomena of phase transitions.

The universal DM density profile in Galaxies has the scaling property: 
\be\label{perfil}
\rho(r) = \rho_0 \; F\left(\displaystyle\frac{r}{r_0}\right) \quad , \quad  
F(0) = 1 \quad , \quad x \equiv  \displaystyle\frac{r}{r_0} \; ,
\ee
where $ r_0 $ is the DM core radius.
As empirical form of cored profiles one can take Burkert's form for $ F(x) $.
Cored profiles {\bf do reproduce} the astronomical observations
(see contributions here by van Eymeren, Gentile and Salucci).

\bigskip

The surface density for dark matter (DM) halos 
and for luminous matter galaxies is defined as: 
$ \mu_{0 D} \equiv r_0 \; \rho_0, $ 
$ r_0 = $ halo core radius, $ \rho_0 = $ central density for DM galaxies.
For luminous galaxies  $ \rho_0 = \rho(r_0) $
(Donato et al. 09, Gentile et al. 09).

Observations show an Universal value for $ \mu_{0  D} $: independent of 
the galaxy luminosity for a large number of galactic systems 
(spirals, dwarf irregular and spheroidals, elliptics) 
spanning over $14$ magnitudes in luminosity and of different Hubble types.
Observed values:
$$
\mu_{0  D} \simeq 120 \; \frac{M_{\odot}}{{\rm pc}^2} = 
5500 \; ({\rm MeV})^3 = (17.6 \; {\rm Mev})^3 \quad , \quad
5 {\rm kpc}  < r_0 <  100 {\rm kpc} \; .
$$
Similar values $ \mu_{0  D} \simeq 80  \; \frac{M_{\odot}}{{\rm pc}^2} $ are observed in 
interstellar molecular clouds of size $ r_0 $ of different type and composition over 
scales $ 0.001 \, {\rm pc} < r_0 < 100 $ pc (Larson laws, 1981).
Notice that the surface gravity acceleration is 
given by $\mu_{0  D}$ times Newton's constant.

{\vskip0.2cm} 

The scaling form eq.(\ref{perfil}) of the density profiles
implies scaling properties for the energy and entropy.

The total energy becomes using the  virial theorem and the profile $F(x)$:
$$
E = \frac12 \; {\avg U} = - \frac14 \; G \; \int 
\frac{d^3r \; d^3r'}{|\rv-\rv'|} \; \avg{\rho(r) \; \rho(r')}=
- \frac14 \; G \; \rho_0^2 \; r_0^5 \int \frac{d^3x \; d^3x'}{|\vx-\vx'|} \; 
\avg{F(x) \; F(x')} \quad \Rightarrow \quad E \sim G \; \mu_{0  D}^2 \; r_0^3
$$ 
Therefore, the energy scales as the volume.

{\vskip0.1cm} 

The Boltzmann-Vlasov distribution function $ f(\pv,\rv) $ for consistency 
with the profile form eq.(\ref{perfil}), must scale as
$$
f(\pv,\rv) = \frac1{m^4 \; r_0^3 \; G^{\frac32} \; \sqrt{\rho_0}} \; 
{\cal F} \left(\frac{\pv}{m \; r_0 \; \sqrt{G \; \rho_0}},
\frac{\rv}{r_0}\right)
$$
where $ m $ is the DM particle mass. Hence, the entropy scales as 
$$
S_{gal} = \int f(\pv,\rv) \; \log f(\pv,\rv) \; d^3 p \;  d^3 r
\sim r_0^3 \; \frac{\rho_0}{m} = r_0^2 \; \frac{\mu_{0 D}}{m}
$$
The {\bf entropy} scales as the {\bf surface} as it is the case for black-holes.
However, for black-holes of mass $ M $ and area $ A = 16 \, \pi \; G^2 \; M^2 $,
the entropy $ S_{BH} = A /(4 \; G) = 4 \, \pi \; G \; M^2 $.
That is, the proportionality coefficients $ c $ between entropy and area are
 very different:
$$
c_{gal}=\frac{S_{gal}}{r_0^2} \sim \frac{\mu_{0 D}}{m} \quad , \quad
c_{BH}= \frac{S_{BH}}{A} = \frac1{4 \; G} \quad {\rm which ~ implies} \quad
\frac{c_{BH}}{c_{gal}} \sim \frac{m}{\rm keV} \; 10^{36}
$$
showing that the entropy per unit area of the galaxy is much smaller 
than the entropy of a black-hole. In other words the Bekenstein 
bound for the entropy of physical is well satisfied here.

\medskip

In order to compute the surface density and the density profiles from
first principles we have evolved the linearized Boltzmann-Vlasov equation 
since the end of inflation till today [2-3].

\begin{figure}
\begin{turn}{-90}
\psfrag{"dengifd.dat"}{}
\psfrag{"dengibe.dat"}{}
\psfrag{"r2dengimb.dat"}{}
\includegraphics[height=9.cm,width=3.5cm]{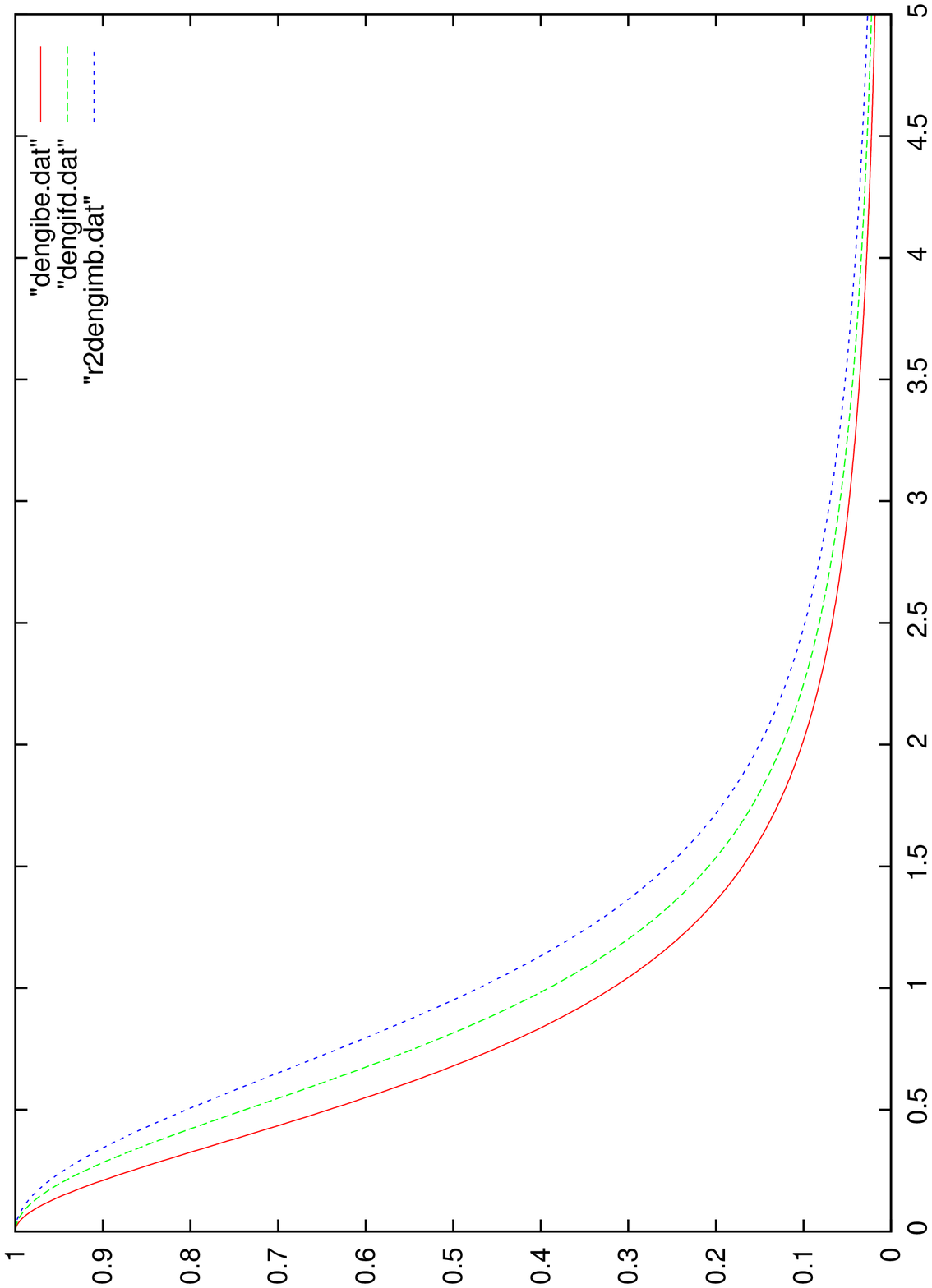}
\end{turn}
\caption{}
\label{perflin}
\end{figure}

We depict in fig. \ref{perflin} the density profiles vs. $ x \equiv r/r_{lin} $
for fermions (green) and bosons (red) decoupling ultrarelativistic
 and for particles decoupling non-relativistically (blue).
These profiles turn to be universal with the same shape as, for example, the Burkert
profile.  $ r_{lin} \sim r_0 $ depends on the galaxy and is entirely determined
theoretically in terms of cosmological parameters and $ m $ [2-3].

\medskip

We obtain in refs. [3,4] for the galaxy surface density in the linear
approximation
$$ 
\mu_{0 \, lin} = 8261 \; \left[\frac{Q_{prim}}{({\rm keV})^4}\right]^{0.161}
\left[1+0.0489 \; \ln \frac{Q_{prim}}{({\rm keV})^4} \right] {\rm MeV}^3 
$$
where $ 0.161 = n_s/6 , \; n_s $ is the primordial spectral index
and fermions decoupling UR were considered.
Matching the {\bf observed values} from spiral galaxies $ \mu_{0 \, obs} $ 
with this $ \mu_{0 \, lin} $ gives the primordial phase-space density
$ Q_{prim}/({\rm keV})^4 $ and 
from it the mass of the DM particle. 
We obtain $ 1.6 < m < 2 $ keV for the dark matter particle mass [4].

\medskip

Linear density profiles turn to be cored at scales $ r \ll r_{lin} $.
At intermediate regime $ r \gtrsim r_{lin} $ we obtain [4],
$$
\rho_{lin}(r)\buildrel{r \gtrsim r_{lin}}\over= 
 \left(\frac{36.45 \; {\rm kpc}}{r}\right)^{1+n_s/2} \; 
\ln\left(\frac{7.932\; {\rm Mpc}}{r}\right) \times 
\left[ 1 + 0.2416 \; \ln \left(\frac{m}{\rm keV}\right)\right] \; 
10^{-26} \;  \frac{\rm g}{{\rm cm}^3} \quad , \quad 1+n_s/2 = 1.482
$$
The theoretical linear results {\bf agree} with the universal empirical behaviour 
$ r^{-1.6\pm 0.4} $: M. G. Walker et al.  (2009) (observations), 
I. M. Vass et al. (2009) (simulations).

\medskip

We summarize in the Table the values for non-universal galaxy quantities 
from the observations and from the linear theory results.
The  larger and less denser are the galaxies, the better are the results from the linear
theory for non-universal quantities. The linear approximation turns to improve for 
larger galaxies (i. e. more diluted) [4]. Therefore, universal quantities as profiles 
and surface density are reproduced by the linear approximation.
The agreement between the linear theory and the observations is {\bf remarkable}.

The last column of the Table corresponds to 100 GeV mass wimps.
The wimps values strongly disagree by several orders of magnitude with the observations.

\medskip

\begin{table*}
 \centering
\begin{tabular}{|c|c|c|c|} \hline  
      & Observed Values & Linear Theory & Wimps in linear theory \\
\hline 
  $ r_0 $ & $ 5 $ to $ 52 $ kpc &  $ 46 $ to $ 59 $ kpc & $ 0.045 $ pc \\
\hline 
  $ \rho_0 $ & $ 1.57  $ to $ 19.3 \times 10^{-25}  \; \frac{\rm g}{{\rm cm}^3} $  & 
$ 1.49  $ to $ 1.91  \times 10^{-25}  \; \frac{\rm g}{{\rm cm}^3} $  &
$ 0.73  \times  10^{-14}  \; \frac{\rm g}{{\rm cm}^3} $ \\ \hline 
  $ \sqrt{{\overline {v^2}}}_{halo} $ & $ 79.3 $ to $ 261 $ \; km/sec & 
$ 260 $ \; km/sec &  $ 0.243 $ \; km/sec \\
\hline   
\end{tabular}
\end{table*}

{\bf References}

\begin{description}
\item[1]  H. J. de Vega, N. G. Sanchez,  arXiv:0901.0922, 
Mon. Not. R. Astron. Soc. 404, 885 (2010).
\vspace{-0.3cm}
\item[2] D. Boyanovsky, H. J. de Vega, N. G. Sanchez, 	
arXiv:0710.5180, Phys. Rev. {\bf D 77}, 043518 (2008).
\vspace{-0.3cm}
\item[3] H. J. de Vega, N. G. Sanchez, arXiv:0907.0006.
\vspace{-0.3cm}
\item[4] H. J. de Vega, P. Salucci, N. G. Sanchez, arXiv:1004.1908.
\end{description}

\newpage

\subsection{Pasquale D. Serpico}

\begin{center}

Physics Department, Theory Group, CERN, CH--1211 Geneva 23, Switzerland and
LAPTH, UMR 5108, 9 chemin de Bellevue - BP 110, 
74941 Annecy-Le-Vieux, France.

\end{center}

\def\Odm{\Omega_{\rm DM}}
\def\mdm{m_{\rm DM}}
\def\ndm{n_{\rm DM}}
\def\edm{\eta_{\rm DM}}
\def\eb{\eta_{\rm b}}
\def\Ob{\Omega_{\rm b}}
\def\d{{\rm d}}
\hyphenation{brems-strah-lung}

\bigskip

\begin{center}

{\bf Astrophysical interpretations of PAMELA/Fermi cosmic ray lepton data}

\end{center}

\medskip

The measurement of the positron fraction in the cosmic ray spectrum  
by the PAMELA satellite collaboration~(Adriani:2008zr) has stimulated 
a lot of theoretical and phenomenological activity (for a slightly more 
extended review see~(Serpico:1900zz)). 
The data show a rise between $\sim 7$ GeV and  100 GeV with a power-law 
index of about $\sim 0.35$, a feature that many have interpreted  as an 
indirect signal  of WIMP dark matter (DM) annihilations/decays.
It is true that, on very general grounds, this behavior is at odds with 
the standard predictions for secondary positrons produced in the collisions 
of cosmic ray nuclei propagating in the inter-stellar medium (ISM),
for which a decreasing fraction with $E$ is expected  (e.g.~(Moskalenko:1997gh)). 
Barring {\it major} flaws in our understanding of cosmic ray astrophysics, 
it appears  that when combined with information on the  spectral 
slope for the overall $e^{+}+e^{-}$ flux $\sim E^{-3}$,  an additional 
source of positrons is required~(Serpico:2008te). In particular, 
this conclusion can be hardly avoided after the publication of the quite hard 
spectrum of $e^{+}+e^{-}$ measured by the Fermi satellite~(fermiel).

Besides DM model building activity, the interest of these data has triggered a 
significant theoretical effort in revisiting the issue of {\it if}
and {\it how plausibly} astrophysical sources can account for the observations, 
as well as possible further tests of these ideas. Although some models exist 
involving a relatively special astrophysical event in the neighborhood of the 
Earth in the recent past,  by far the most popular astrophysical explanations 
invoke a whole class (or mechanism) of leptonic accelerators.
Not surprisingly, virtually all models of this kind attribute ultimately the 
powering source to the death of massive stars as supernovae (SNe). 

After the explosion of a SN,  about $\sim 10^{51}$ erg of  kinetic energy  are 
released to the outwards-moving ejecta, powering (for thousands of years) 
shell-like SN Remnants (SNRs)~(Reynolds08). The first-order, non-relativistic 
shock acceleration when the SNR expands in the ISM is the standard paradigm for the 
acceleration of both $e^{-}$ and protons/nuclei, with energies from GeV to 
PeV~(Blandford:1987pw). 

In this framework, several possibilities have been explored  in the literature. 
For example, one is that hard spectra of high energy $e^{+}$ and $e^{-}$ are 
produced as secondaries in  polar caps of SN whose progenitors were massive 
magnetic stars with winds, like Wolf Rayet and Red Super Giants~(Biermann:2009qi). 
While energetics and spectral index requirements can be reproduced, it is 
interesting to ask the even more basic question of if peculiar field configurations 
and very massive stars are in fact needed at all, i.e. what are the 
predictions of the ``baseline'' model of SNRs. In fact, it was argued 
in~(Blasi:2009hv) that already internal production and reacceleration in magnetic
field configurations parallel to the normal to the shock front might be enough.
In this model, the `excess' is interpreted as the additional component due to 
positrons created as byproducts of hadronic interactions inside old SNRs, the 
standard source of the bulk of sub-TeV cosmic rays. Naively, one would expect 
internal production to be responsible for $e^{+}$ having a spectrum as hard as 
the proton one, which would translate into a flat $e^+$ fraction emerging at 
hundreds of GeV. However, if the environment has a diffusion coefficient $D(E)$ 
growing with energy (as expected), {\it reacceleration} involves secondary
 particles which are produced within a distance proportional to $D(E)$ from
the shock (on both sides): the higher the energy, the more secondaries are 
involved in the process,
which can in principle produce spectra harder than primary ones, although its 
quantitative details depend on environmental conditions during the late stages 
of evolution of SNRs. So, it is ultimately observations which should determine 
how relevant this process is in nature.  Fortunately, due to the hadronic nature 
of the process, spallation unavoidably produces a large flux of antiprotons 
at $E> 100\,$GeV~(Blasi:2009bd) as well as signatures in secondary to 
primary nuclei~(Mertsch:2009ph).

In most of the cases where a core-collapse SN takes place, 
however, one expects another ``relic'': a highly magnetized, 
fast spinning neutron star. The pulsar loses energy steadily by 
spinning-down, and this energy is transferred to electromagnetic 
channels, with some fraction going into relativistic electron-positron pairs. 
 The acceleration of leptons stripped from the pulsar and the generation
of a very high-energetic electromagnetic cascade is not 
hard to achieve. The spectral shape of the
emission is however determined by a complicated series of plasma processes, 
related also to the location of the acceleration: $e^{\pm}$ and $\gamma$'s 
are continuously produced, slowed-down and reconverted into each other by 
curvature radiation, inverse-compton scattering, photon-photon pair creation, etc. 
In turn, the e$^\pm$ pair wind of 
relativistic particles leaving the magnetosphere forms the so-called Pulsar 
Wind Nebula (PWN)~(Gaensler:2006ua). For several pulsars, observations 
ranging from radio to X-ray data suggest that, at the termination shock
 produced when the PW interacts with the surrounding slower ejecta, 
efficient acceleration takes place, resulting into quite hard lepton spectra ($\sim E^{-1.5}$) extending up to hundreds of GeV. If a non-negligible fraction of this population makes its way out to the ISM, it is possible to reproduce quite effectively the data~(pulsars).
Compared with explanations invoking SNRs, pulsars have the big advantage of being known to host a {\it large} population of non-thermal, relativistic pairs. On the other hand, detailed predictions of the spectral shape are rather uncertain and the theory of {\it relativistic} shocks in non-trivial magnetic field configurations and a medium containing pairs (possibly ``polluted'' by some protons/nuclei) is less well understood. 
Studies exist, however, where high efficiencies and hard spectra have been found, see e.g.~(Amato:2006ts).

It is clear by now that the once-popular ``standard model'' of secondary cosmic ray positrons is insufficient in describing the data {\it at high energy}. However, it should be recognized that models relying only on ISM productions had been selected in the past due to their simplicity, 
rather than for the lack of astrophysical sources  proposed to contribute to lepton fluxes (an example of 15 years-old study in this sense is~(Atoian:1995ux)). These sources are nothing but the same objects shining in the sky in X and gamma-ray band. It should go without saying that they are {\it naturally} the first class of objects to look at for explaining unaccounted observations of cosmic ray fluxes. It is actually a sign of progress of the field that, thanks to observational improvements, it is now time to include them in theoretical models and fits. Concerning the implications for  the WIMP paradigm for DM, the present data by themselves are rather neutral to it, since typical expectations for DM signals in antimatter fall a couple of orders of magnitude below the fluxes observed. On the other hand it is fair to conclude that, until a better understanding of the astrophysical sources is achieved, most antimatter signatures of WIMPs are far from robust (with the possible exception of sharp spectral edges or ``large'' numbers of low-energy antideuterons.) 

Although there is still a long way to finally settle the interpretation of the data, the obvious next step would be to identify if the {\it main} contributors to the observed fluxes are leptonic accelerators (such as pulsars) or hadronic ones (such as SNRs). This might be a task for the forthcoming  AMS-02 mission~(AMS-02).

\bigskip

{\bf References}

\medskip

\begin{itemize}
\item{Adriani:2008zr}
O.~Adriani {\it et al.}  [PAMELA Collaboration],
    Nature 458, 607-609, 2009.
\item{Serpico:1900zz}
  P.~D.~Serpico,
  Nucl.\ Phys.\ Proc.\ Suppl.\  {\bf 194}, 145 (2009).
\item{Moskalenko:1997gh}
  I.~V.~Moskalenko and A.~W.~Strong,
  Astrophys.\ J.\  {\bf 493}, 694 (1998).
\item{Serpico:2008te}
  P.~D.~Serpico,
  Phys.\ Rev.\  D {\bf 79}, 021302 (2009).
\item{fermiel}
A. A. Abdo {\it et al.} [FERMI-LAT Collaboration],
 Phys. Rev. Lett. {\bf 102}, 181101 (2009).
\item{Reynolds08}
S.P.~Reynolds,
Ann.\ Rev.\ Astron.\ Astrophys.\  {\bf 46}, 89 (2008).
\item{Blandford:1987pw}
  R.~Blandford and D.~Eichler,
  Phys.\ Rept.\  {\bf 154}, 1 (1987).
\item{Biermann:2009qi}
  P.~L.~Biermann, J.~K.~Becker, A.~Meli, W.~Rhode, E.~S.~Seo and T.~Stanev,
  Phys.\ Rev.\ Lett.\  {\bf 103}, 061101 (2009).
\item{Blasi:2009hv}
  P.~Blasi,
  Phys.\ Rev.\ Lett.\  {\bf 103}, 051104 (2009).
\item{Blasi:2009bd}
  P.~Blasi and P.~D.~Serpico,
  Phys.\ Rev.\ Lett.\  {\bf 103}, 081103 (2009).
\item{Mertsch:2009ph}
  P.~Mertsch and S.~Sarkar,
  Phys.\ Rev.\ Lett.\  {\bf 103}, 081104 (2009).
\item{Gaensler:2006ua}
  B.~M.~Gaensler and P.~O.~Slane,
  Ann.\ Rev.\ Astron.\ Astrophys.\  {\bf 44}, 17 (2006).
\item{pulsars}
  D.~Hooper, P.~Blasi and P.~D.~Serpico,
  JCAP {\bf 0901}, 025 (2009);
  H.~Yuksel, M.~D.~Kistler and T.~Stanev,
Phys.\ Rev.\ Lett.\  {\bf 103}, 051101 (2009);
  S.~Profumo,
  arXiv:0812.4457;
  D.~Malyshev, I.~Cholis and J.~Gelfand,
  Phys.\ Rev.\  D {\bf 80}, 063005 (2009);
  N.~Kawanaka, K.~Ioka and M.~M.~Nojiri,
  Astrophys.\ J.\  {\bf 710}, 958 (2010);
    D.~Grasso {\it et al.} [FERMI-LAT Collaboration],
    Astropart.\ Phys.\  {\bf 32}, 140 (2009);
   L.~Gendelev, S.~Profumo and M.~Dormody,
  JCAP {\bf 1002}, 016 (2010);
  T.~Delahaye, J.~Lavalle, R.~Lineros, F.~Donato and N.~Fornengo,
  arXiv:1002.1910.
\item{Amato:2006ts}
  E.~Amato and J.~Arons,
  Astrophys.\ J.\  {\bf 653}, 325 (2006).
\item{Atoian:1995ux}
  A.~M.~Atoian, F.~A.~Aharonian and H.~J.~Volk,
  Phys.\ Rev.\  D {\bf 52}, 3265 (1995).
\item{AMS-02}\texttt{http://ams.cern.ch/}
\end{itemize}

\newpage

\subsection{Rainer Stiele}

\begin{center}

\vskip -0.3cm

and

\medskip

{\bf Tillmann Boeckel, J\"urgen Schaffner-Bielich}, 

\medskip

Institute for Theoretical Physics, Heidelberg University, 
Philosophenweg 16, D-69120 Heidelberg, Germany

\bigskip

{\bf Cosmological bounds on dark matter self-interactions}

\bigskip

R.Stiele@ThPhys.Uni-Heidelberg.de

\end{center}

\def\aap{A\&A}%
\def\apj{ApJ}%
\def\apjl{ApJ Letters}%
\def\apjs{ApJS}%
\def\ijmpa{IJMPA}%
\def\jcap{JCAP}%
\def\mnras{MNRAS}%
\def\physrep{Physics Reports}%
\def\prl{Phys.~Rev.~Lett.}%
\def\ptps{PTPS}%
\newcommand{\mrm}[1]{\mathrm{#1}}

Finite elastic dark matter self-interactions can be one of the 
ingredients to solve the puzzles of collisionless cold dark matter. 
In addition to a summary of the effects on dark matter halos, 
we present cosmological implications of a dark matter 
self-interaction energy density.

\bigskip

While the existence of dark matter (DM) is beyond all question, its 
properties are still subject to debate. Open issues are among others the 
number of galactic substructures and the cusp vs.\ core problem. Observational 
results on and attempts to resolve these shortcomings of collisionless cold DM 
(CDM) are presented in other lectures of this workshop. An additional effect 
related to these topics can be caused by finite DM elastic self-interactions 
(Spergel2000) which we will study in the following.

Substructures are colder halos embedded in larger hotter ones. 
Hence, they become heated and destroyed either by spallation or 
evaporation (Wandelt2000). 
But galactic halos as substructure of clusters have to survive at least for a 
Hubble time (Gnedin2001). Energy transfer occurs also from the hotter outer regions 
to the colder center of halos, so initial cusps are transformed to cores. To explain the 
observed core sizes of less dense dwarf galaxies one requires a larger cross section than 
compatible with cluster core sizes. A velocity dependent cross section can solve this 
dilemma (Dave2001, Yoshida2000). An additional effect of the isotropization of the 
velocity distribution in dense regions is the formation of spherical centers 
(Miralda2002). Cluster collisions allow to constrain the self-scattering strength by 
measuring the offset between the mass peak and the galaxy distribution and the mass to light 
ratio of the collided clusters (Randall2008).
\begin{table}[h]
 \caption{Bounds on CDM self-scattering cross section from halo properties.}
 \begin{tabular}{c|c|c}
 & $\sigma_\mrm{SI}/m_\mrm{DM}\,\left[\mrm{cm}^2/\mrm{g}\right]$ & Ref.\\
 \hline
 Core sizes & $\lesssim 0.5 - 5$  & (Dave2001, Yoshida2000)\\
 Galactic evaporation &  $\lesssim 0.3 $ & (Gnedin2001)\\ 
 Cluster ellipticity & $\lesssim 0.02 $ & (Miralda2002)\\ 
 Bullet cluster & $< 0.7 - 1.25 $ & (Randall2008)
 \end{tabular}
\end{table}

Very recently we highlighted in Ref.\ (Stiele2010) another 
consequence of elastic dark matter self-interactions, namely an additional 
energy density contribution.\\
We describe two particle interactions between scalar bosons ($\phi$) or 
fermions ($\psi$) by minimal coupling to a vector field $V_\mu$. The effective Lagrangian reads:
\begin{subequations}
\begin{eqnarray}
 {\cal L}_\phi&=&{\cal D}_\mu^* \phi^* {\cal D}^\mu \phi - m_\phi^2 \phi^*\phi- 
\tfrac{1}{4} V_{\mu\nu} V^{\mu\nu} + \tfrac{1}{2}m_\mrm{v}^2 V_\mu V^\mu,\ \ \quad\\
 {\cal L}_\psi&=&\bar{\psi}\left(i\slashed{\cal D}-m_\psi\right)\psi- 
\tfrac{1}{4} V_{\mu\nu} V^{\mu\nu} + \tfrac{1}{2}m_\mrm{v}^2 V_\mu V^\mu\;,
\end{eqnarray}
\end{subequations}
with $V_{\mu\nu}=\partial_\mu V_\nu -\partial_\nu V_\mu$ and 
${\cal D}_\mu = \partial_\mu + i g_{\mrm{v} \phi(\psi)} V_\mu$, where 
$g_{\mrm{v}\phi(\psi)}$ is the $\phi(\psi)$-$V$ 
coupling strength. The interaction is repulsive which avoids an enhancement of the annihilation 
cross-section due to the formation of bound states.
The total energy density and pressure of the self-scattering DM can be
 determined from the energy-momentum tensor to be
\begin{equation}
 \varrho_{\phi(\psi)} = \varrho_{\phi(\psi)}^\mrm{free} + 
\frac{g_{\mrm{v}\phi(\psi)}^2}{2 m_\mrm{v}^2} \,n_{\phi(\psi)}^2\;.
\end{equation}
We denote the particle masses $m_\mrm{SIDM}\equiv{m_{\phi(\psi)}}$, 
$m_\mrm{SI}\equiv{m_\mrm{v}}$, and define the coupling constant 
$\alpha_\mrm{SI}\equiv{g_\mrm{v\phi(\psi)}^2/2}$, so that the 
energy density and pressure contributions from DM self-interactions read
\begin{equation}
 \varrho_\mrm{SI}=\frac{\alpha_\mrm{SI}}{m_\mrm{SI}^2}\,n_\mrm{SIDM}^2=p_\mrm{SI}\;,
 \label{eq:eos_SI}
\end{equation}
with $m_\mrm{SI}/\!\sqrt{\alpha_\mrm{SI}}$ as the energy scale 
of the self-interaction. For weak interactions the interaction strength is 
$m_\mrm{weak}/\!\sqrt{\alpha_\mrm{weak}}\sim300\,\mrm{GeV}$ and for 
strong interactions $m_\mrm{strong}/\!\sqrt{\alpha_\mrm{strong}}\sim100\,\mrm{MeV}$.\\
Refs.\ (Narain2006, Agnihotri2009) explored implications 
of this self-interaction energy density contribution on the mass-radius 
relation of compact stars made of self-interacting DM.\\
The equation of state (\ref{eq:eos_SI}) as input to the Friedmann equations 
determines the scaling behaviour of the self-interaction energy density with the 
scale factor $a$:
\begin{equation}
 \varrho_\mrm{SI}\propto{a^{-6}}\;.
 \label{eq:rhoSI_a}
\end{equation}
Hence, $\varrho_\mrm{SI}$ shows the steepest decrease 
and the universe could be in a self-interaction dominated epoch prior to 
radiation domination in the very early universe.\\
Eqs.\ (\ref{eq:eos_SI}) and (\ref{eq:rhoSI_a}) imply that $n_\mrm{SIDM}\propto{a^{-3}}$. 
So the self-interacting DM particles have to be warm for decoupling during 
self-interaction domination and can only be cold if they decouple after the
 self-interaction dominated era, which can last at the latest until shortly 
before primordial nucleosynthesis (see below). In any case, the recently 
highlighted DM particle mass scale in the keV range (deVega2010, Boyanovsky2008) 
is an attractive particle candidate for self-interacting DM.\\

Primordial nucleosynthesis (BBN) is the physical process of choice to 
constrain the self-interaction strength, since the element abundances 
are sensitive to the energy content of the universe via the expansion rate 
$H\propto{\varrho^{1/2}}$. 
The relative contribution of the self-interaction energy density is largest 
at the earliest stage of BBN, the freeze-out of the neutron to proton number 
ratio.	Nearly all neutrons available for the nucleosynthesis processes are 
incorporated into $^4\mrm{He}$, so we can translate the upper limit on the 
primordial $^4\mrm{He}$ abundance from observations $Y_\mrm{P}<0.255$ 
(2$\sigma$, (Steigman2007)) into the following constraint on the 
DM self-interaction strength (Stiele2010):
\begin{equation}
 \frac{m_\mrm{SI}}{\sqrt{\alpha_\mrm{SI}}}\gtrsim1.70\,\mrm{keV}\times
\frac{F_\mrm{SIDM}^0}{m_\mrm{SIDM}/1\,\mrm{keV}}\;.
\end{equation}
The relative amount of SIDM $F_\mrm{SIDM}^0\equiv\Omega_\mrm{SIDM}^0/\Omega_\mrm{DM}^0$ 
serves to include the possibility of multiple DM components.\\
Even an additional energy density contribution of DM self-interactions of the 
strength of the strong interaction 
($m_\mrm{strong}/\!\sqrt{\alpha_\mrm{strong}}\sim100\,\mrm{MeV}$) 
is consistent with the primordial element abundances.

Which further consequences can a self-interaction dominated universe before 
BBN have? A physical process happening during this early stage is the decoupling 
of the DM particles. Chemical decoupling occurs when the expansion rate of the 
universe exceeds the DM annihilation rate $\Gamma_\mrm{A}=n_\mrm{DM}\,
\langle{\sigma_\mrm{A}}v\rangle$. In a universe that is dominated by a 
WDM self-interaction energy density contribution, also the expansion rate 
$H\propto{\varrho^{1/2}}$ is proportional to the WDM particle density , 
so that  the WDM annihilation cross-section is independent on the particle 
parameters but determined by the elastic self-interaction strength:
\begin{equation}
 \sigma_\mrm{A}^\mrm{WDM}\approx7.45\times10^{-7}\times
\frac{100\,\mrm{MeV}}{m_\mrm{SI}/\!\sqrt{\alpha_\mrm{SI}}}\,\sigma_\mrm{weak}\;,
\end{equation}
with $\sigma_\mrm{weak}\approx1.24\times10^{-39}\,\mrm{cm^2}$.
Hence, WSIDM decoupling in a self-interaction dominated universe reproduces 
naturally and consistently the {\textquoteleft}super weak' inelastic coupling 
between the WSIDM and baryonic matter.\\
For decoupling of a collisionless CDM component, usually represented by WIMPs, 
in a self-interaction dominated universe the natural scale of the velocity 
weighted mean annihilation cross-section becomes (Stiele2010)
\begin{eqnarray}
 \langle\sigma_\mrm{A}v\rangle_\mrm{CDM}&\approx&2.77\times10^{-23}\,\mrm{cm^3\,s^{-1}} \\
 &&\times\,\frac{m_\mrm{CDM}/10\,\mrm{TeV}}{m_\mrm{WDM}/1\,\mrm{keV}}\,
\frac{1\,\mrm{MeV}}{m_\mrm{SI}/\!\sqrt{\alpha_\mrm{SI}}}\,
\frac{F_\mrm{WDM}^0}{1-F_\mrm{WDM}^0}\;. \nonumber
\end{eqnarray}
Hence, the natural scale of the CDM annihilation cross-section depends 
beside the WDM elastic self-interaction strength also linearly on the 
CDM particle mass. All in all the natural scale of CDM decoupling can be 
increased by some orders of magnitude. This is in contrast to the 
\textquoteleft{WIMP miracle}' (meaning that they actually aren't so 
weakly interacting) and interestingly enough, such boosted CDM 
annihilation cross-sections are able to explain the high energy 
cosmic-ray electron-plus-positron spectrum measured by Fermi-LAT 
and the excess in the PAMELA data on the positron fraction (e.g.\ (Bergstroem2009B)).

Another consequence of an early self-interaction dominated epoch may concern structure 
formation. A relativistic analysis of ideal fluid cosmological perturbations reveals 
for  self-interaction dominated DM the following evolution of the density contrast 
$\delta$ in the subhorizon limit (${k_\mrm{ph}}/{H} \gg 1$):
\begin{equation}
\delta_{\mrm{SIDM}} \propto a \cdot \left(A \cos(a^2 - 3\pi/4) + B \sin(a^2 - 3\pi/4)\right)\;,
\end{equation}
i.e.\ an oscillation with linearly growing amplitude (Hwang1993). However, 
any increase in the density contrast in WSIDM produced will be washed out either 
due to collisional self-damping or due to free streaming.\\
But a subdominant collisionless CDM component allows some increase in density 
fluctuations to be stored (Stiele2010):
\begin{equation}
 \delta_{\mrm{CDM}} = a \cdot \left(C/{a^\mrm{in}_{k}}^2\right) + D\;.
\end{equation}
This means that subhorizon collisionless CDM density fluctuations will also grow 
linearly during a self-interaction dominated phase. Thus fluctuations at low 
masses in the matter power spectrum are enhanced. They are limited by the 
comoving wavenumber that is equal to the Hubble scale at self-interaction--radiation
 equality, coresponding to $\sim1.4 \times 10^{-3} M_\odot$ as the largest 
structures that can be affected (Stiele2010).

Another physical process in the early universe, 
maybe influenced if happening 
during a self-interaction dominated epoch is the QCD phase transition, 
which 
nature has recently received new attention (Boeckel2009).

This work was supported by the German Research Foundation (DFG) 
through the 
Heidelberg Graduate School of Fundamental Physics 
(HGSFP) and the Graduate Program for 
Hadron and Ion Research by the Gesellschaft f\"ur 
Schwerionenforschung (GSI), Darmstadt.

\bigskip

{\bf References}

\medskip

\begin{itemize}
\item{Spergel2000}
Spergel, D.~N. and others, PhysRevLett.84.3760 (2000).
\item{Wandelt2000}
Wandelt, Benjamin D. and others, astro-ph/0006344
\item{Gnedin2001}
Gnedin, O.~Y. and others, 2001ApJ...561...61G
\item{Dave2001}
Dav{\'e}, R. and others, 2001ApJ...547..574D
\item{Yoshida2000}
Yoshida, N. and others, 2000ApJ...544L..87Y
\item{Miralda2002}
Miralda-Escud\'e, J., 2002ApJ...564...60M
\item{Randall2008}
Randall, S.~W. and others,2008ApJ...679.1173R
\item{Stiele2010}
Stiele, R. and others, PhysRevD.81.123513 (2010)
\item{Narain2006}
Narain, G. and others, PhysRevD.74.063003 (2006)
\item{Agnihotri2009}
    Agnihotri, P. and others, PhysRevD.79.084033 (2009)
\item{Boeckel2007}
Boeckel, T. and others, PhysRevD.76.103509 (2007)
\item{deVega2010}
de Vega, H.~J. and others, 2010MNRAS.404..885D 
\item{Boyanovsky2008}
Boyanovsky, D. and others, PhysRevD.77.043518 (2008)
 \item{Steigman2007}
Steigman, G., annurev.nucl.56.080805.140437  (2007)
\item{Bergstroem2009B}
Bergstr{\"o}m, L. and others, PhysRevLett.103.031103 (2009)
\item{Hwang1993}
Hwang, J.-C., 1993ApJ...415..486H
\item{Boeckel2009}
Boeckel, T. and others, arXiv0906.4520
\item{Boeckel2010}
Boeckel, T. and others, arXiv1002.1793
\end{itemize}

\newpage

\subsection{Janine van Eymeren}

\vskip -0.2cm

\centerline{ Universit\"at Duisburg-Essen, 
The University of Manchester.}

\newcommand{\HI}{H\,{\sc i}}
\newcommand{\skms}{\ensuremath{\,\mbox{km}\,\mbox{s}^{-1}}}

\bigskip

\bigskip

\begin{center}

{\bf Non-circular motions and the cusp-core discrepancy in dwarf galaxies}

\end{center}

\medskip

The cusp-core discrepancy, one of the major problems in galaxy evolution, 
is still causing many debates between observers and cosmologists. Cold Dark Matter (CDM) 
simulations predict cuspy haloes in the central parts of galaxies with a 
density distribution 
described by a power law $\rho(r)\sim r^{\alpha}$ with $\alpha$ ranging from -1 (e.g., [7]) 
to -1.5 (e.g., [6]). This cusp leads to a steeply rising rotation curve. However, 
observations of dark matter dominated dwarf and low surface brightness (LSB) galaxies 
show that their rotation curves rise less steeply than predicted by 
CDM simulations (e.g., [1]). 
At small radii (typically a few kpcs), the mass distribution can better 
be described by a central, 
constant-density core (e.g., [4]). Observational shortcomings like the effect of 
beam smearing or 
slit misplacement have been claimed to be responsible for flattening the rotation 
curves in the 
inner 1\,kpc. Furthermore, deviations from the assumed circular orbits, so-called 
non-circular 
motions might also lead to an underestimate of the slope.

\medskip

We chose a sample of six nearby irregular dwarf galaxies and obtained \HI\ synthesis 
observations with sufficiently high (i.e., below 1\,kpc) spatial resolution. 
We derived the kinematic parameters by performing a tilted-ring analysis of the 
Hermite velocity fields. In order to quantify the contribution of non-circular 
motions to the derived rotation curves, we performed a harmonic decomposition up to 
third order. After that, the rotation curves were decomposed into contributions from 
gas, stars, and dark matter, where the distribution of the dark matter was fitted with 
both NFW and pseudo-isothermal (ISO) halo profiles. We then performed a $\chi^2$ 
minimisation to find the best fit.

\medskip

\begin{figure}
\centering
\includegraphics[width=\textwidth]{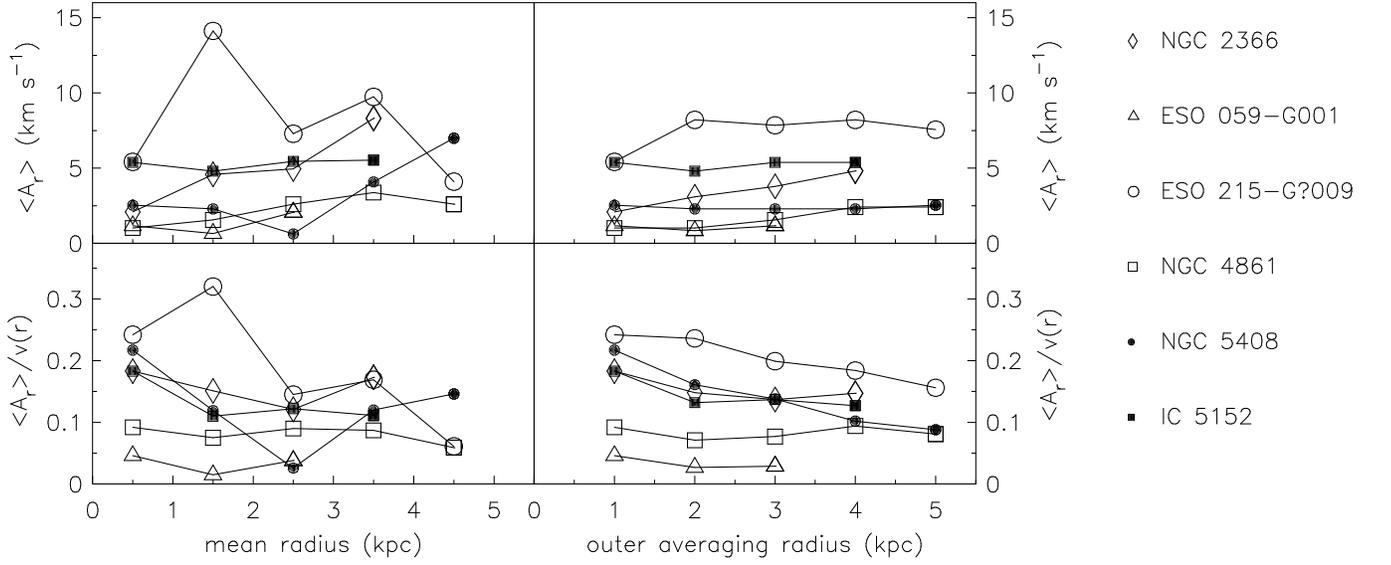}
\caption{{\bf Upper left panel:} the mean values of the quadratically-added 
amplitudes of the non-circular motions within rings of 1\,kpc width (i.e., $0<r<1$\,kpc, 
$1<r<2$\,kpc, ..., $4<r<5$\,kpc) for each galaxy (indicated by different symbols). 
{\bf Upper right panel:} the same as the upper left panel, but the amplitudes of the 
non-circular motions are averaged within rings of increasing radius 
(i.e., $0<r<1$\,kpc, $0<r<2$\,kpc, ..., $0<r<5$\,kpc). {\bf Lower left and right panel:} 
like the upper left and right panel, but the amplitudes of the non-circular motions 
are normalised by the local rotation velocity.}
\label{Fignoncirc}
\end{figure}

Figure~\ref{Fignoncirc} shows the quadratically-added amplitudes of the non-circular 
components up to third order \emph{vs.} the distance from the dynamic centre (upper row). 
All sample galaxies have non-circular motions with absolute amplitudes below 10\skms, 
often below 5\skms, independent of the radius we look at. The lower row of 
Fig.\ref{Fignoncirc} shows the quadratically-added amplitudes normalised by the local 
rotation velocity. According to simulations by [5] (and ref. therein), non-circular 
motions add up to about 50\% of the local rotation velocity at a radius of 1\,kpc and 
even more below 1\,kpc. Here, it can be seen that the non-circular motions contribute 
less than 25\% to the local rotation velocity (again independent of the radius). 
This means that the non-circular motions in our sample galaxies are not high enough 
to significantly change the slope of the rotation curves.  

\medskip

\begin{SCfigure}
\centering
\includegraphics[width=.5\textwidth]{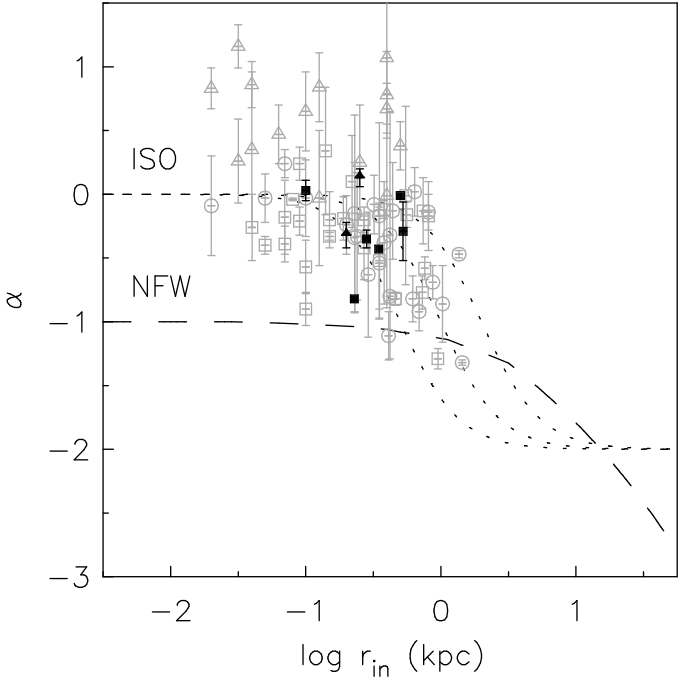}
\caption{The inner slope $\alpha$ of the dark matter density profiles plotted against
  the radius of the innermost point. Grey symbols and black triangles present 
observations of dwarf and LSB galaxies performed by different observers (see [3] 
for details). Our results are overplotted with black squares.}
\label{Figalpha}
\end{SCfigure}

The results from the mass decomposition can be summarised as follows: 
$\chi^2_{\rm{red}}$ is in almost all cases much larger when we use the 
NFW model as a representation of the dark matter distribution. 
The adapted parameters of the ISO model are plausible and follow the recent 
finding of a constant central surface density ([2]). We found values for the slope 
$\alpha$ of $-0.43$ up to $0.03$ (see Fig.~\ref{Figalpha}), which is within the 
uncertainties of the value measured from the observations of LSB galaxies, 
$-0.2\pm0.2$ ([1]).

\medskip

We can rule out a significant contribution of non-circular motions in the central 
parts of our sample galaxies. The slopes can better be described by the empirically 
derived ISO halo. Therefore, we conclude that the measured cores are not hidden cusps. 
This result is in agreement with many other publications (Fig.~\ref{Figalpha}). 
For further details see [3].

\bigskip

{\bf References}

\medskip

\begin{description}
\item[{\bf[1]}] de Blok, \& Bosma 2002, A\&A, {\bf 385}, 816
\vspace{-0.2cm}
\item[{\bf[2]}] Donato et al. 2009, MNRAS, {\bf 397}, 1169
\vspace{-0.2cm}
\item[{\bf[3]}] van Eymeren, Trachternach, Koribalski, \& Dettmar 2009, A\&A, {\bf 505}, 1
\vspace{-0.2cm}
\item[{\bf[4]}] Gentile, et al. 2004, MNRAS, {\bf 351}, 903
\vspace{-0.2cm}
\item[{\bf[5]}] Hayashi, \& Navarro 2006, MNRAS, {\bf 373}, 1117
\vspace{-0.2cm}
\item[{\bf[6]}] Moore, et al. 1999, MNRAS, {\bf 310}, 1147
\vspace{-0.2cm}
\item[{\bf[7]}] Navarro, Frenk, \& White 1996, ApJ, {\bf 462}, 563
\end{description}

\newpage

{} {} $ \; \; {} $

\newpage

\subsection{Markus Weber}

\vskip -0.2cm

\begin{center}

Institut f\"ur Experimentelle Kernphysik (EKP)\\
  Karlsruher Institut f\"ur Technologie (KIT)\\

\vspace{1.cm}

  {\bf Determination of the local DM density in our Galaxy \\[5mm]}

\end{center}

\vspace{0.5cm}

The best evidence for Dark Matter (DM) in galaxies is usually provided by rotation
velocities, which do not fall off fast enough at large distances from the 
Galactic centre (GC). For a flat rotation curve the DM density has to fall off 
like 1/r$^2$ at large distances, which is indeed expected from the virial 
theorem for a gas of gravitational interacting particles.\\
A reliable determination of the local DM density is of great interest for
direct DM search experiments, where elastic collisions between WIMPs and
the target material of the detector are searched for. This signal is proportional
to the local density.\\ 

The local DM density can be determined from the rotation velocities of
Galactic objects which are proportional to the gravitational potential of the
Milky Way (MW). The gravitational potential is connected to the density
distribution of the Galaxy via Poisson's equation.
The first step of the determination of the local DM density is the
assumption of the density distributions of the two different matter
contributions to the MW: the luminous and the dark matter. Then, the 
parameters of these distributions are fitted to experimental constraints
obtained from Galactic kinematics.

The density distribution of
the luminous matter of a spiral galaxy is split into two parts,
the Galactic disc and the Galactic bulge. The parametrization of
the density distribution of the bulge is adapted from
the publication by (Cardone:2005)
\begin{eqnarray}
  \rho_{b}(r,z) & = & \rho_b \cdot \left( \frac{\tilde{r}}{r_{0,b}}
  \right)^{-\gamma_b} \cdot \left( 1 + \frac{\tilde{r}}{r_{0,b}} \right)^{\gamma_b -
  \beta_b} \exp{\left( -\frac{\tilde{r}^2}{r_t^2}\right)}\\
  \nonumber \tilde{r}^2 & = & \sqrt{x^2+y^2 + (z/q_b)^2}.
\label{eq:bulge}
\end{eqnarray}
The parameters of the Galactic bulge are obtained from the consideration
of the velocity distribution in the Galactic disc near the GC.
The parametrization of the Galactic disc is taken from the publication
by Sparke
\begin{equation}
  \rho_{d}(r,z) = \rho_d \cdot \exp(-r/r_d) \cdot \exp(-z/z_d).
\label{eq:disc}
\end{equation}
The parameter $\rho_\mathrm{d}$ describes the density of the Galactic disc at the
GC while r$_\mathrm{d}$ and z$_\mathrm{d}$ describe the scale parameter in radial and vertical
direction.\\
The Galactic DM contribution is distributed within a 
large DM halo which extends the visible matter by an order
of magnitude. It is commonly believed that the profile of a DM halo can
be well fitted by the universal function
\begin{eqnarray}
  \rho_\chi(r) & = & \rho_{\odot,DM} \cdot \left ( \frac{\tilde r}{r_\odot} \right )^{-\gamma}
  \cdot \left \lbrack \frac{1 + \left ( \frac{\tilde r}{a} \right )^\alpha}{1 + \left
  ( \frac{r_\odot}{a} \right )^\alpha} \right \rbrack^{\frac{\gamma -
  \beta}{\alpha}},\\
  \nonumber \tilde r & = & \sqrt{x^2 + \frac{y^2}{\epsilon_{xy}^2} + \frac{z^2}{\epsilon_z^2}}.
\end{eqnarray}
Here, a is the scale radius of the density profile, which determines
at what distance from the centre the slope of the profile changes,
$\epsilon_{\mathrm{xy}}$ and $\epsilon_{\mathrm{z}}$ are the eccentricities of the
DM halo within and perpendicular to the Galactic plane and r$_\odot$
is the Galactocentric distance of the solar system.

The local DM density is determined from the kinematics of
stars and interstellar gas. The observation of the Galactic kinematics
depends on the standard frame of rest which means the Galactocentric
distance of the Solar System $r_\odot$ and its rotation velocity $v_\odot$.
These values are obtained from the observation of Sgr A* which is
probably the centre of the Galaxy because of its own small velocity.
The distance between the Sun and the GC has been determined to 
Gillessen:2008:
 \begin{equation}
r_\odot= 8.33 \pm 0.35\ \mathrm{kpc},
\label{r0}
\end{equation}
in agreement with previous authors Ghez:2008.
With this Galactocentric distance one finds a rotation velocity of the Sun
\begin{equation}
v_\odot = 244 \pm 10\ \mathrm{km}\ \mathrm{s}^{-1},
\label{vsun}
\end{equation}
which is consistent with recent observations of Galactic masers
in Bovy:2009, who used 
data from the Very Long Baseline Array (VLBA) and the 
Japanese VLBI Exploration of Radio Astronomy (VERA).

The experimental measurements to constrain the local
DM density are i) the total matter density at the Sun, ii) the
total Galactic mass within a Galactocentric distance of
60 kpc, iii) the surface density of the luminous and iv) the total
matter. The velocity distribution in the Galactic disc and
the height of the interstellar gas distribution are used to
constrain the density distribution in radial and 
in z direction.\\
The total matter density at the Sun is determined from
the vertical motion of stars to be 
$\rho_{\odot,\mathrm{tot}} (z=0) = 0.102 \pm 0.010\ \mathrm{M}_\odot\ \mathrm{pc}^{-3}$.
This measurement was first proposed and performed by
Jan Oort in 1932.\\
The surface density of the luminous matter at the Sun
is used to constrain the matter contribution of the Galactic
disc. It is obtained from star counts to be 
$\Sigma_{\mathrm{vis}} = 35 - 58$ M$_\odot$ pc$^{-2}$ (Naab:2005).
The total surface density is determined in the paper by
Holmberg:2004 from the modeling of the vertical gravitational
potential. This analysis resulted in 
$\Sigma (< 1.1\ \mathrm{kpc}) =$ 74 $\pm$ 6 M$_\odot$ pc$^{-2}$.\\ 
The total mass of the MW within a Galactocentric distance of
60 kpc is obtained from the observation of the kinematic of
halo stars. Using a large sample of 2400 blue horizontal-branch stars 
from the Sloan Digital Sky Survey in the
halo (z $>$ 4 kpc, R$<60$ kpc) and comparing the results with
N-body simulations using an NFW profile from Xue:2008 find
\begin{equation}
\mathrm{M}_{\mathrm{R}<60~{\mathrm{kpc}}} =4.0 \pm 0.7 \cdot 10^{11} \mathrm{M}_\odot ,
\label{mtot}
\end{equation}
which corresponds to M$_{\mathrm{tot}} = 1.0^{+0.3}_{-0.2} \cdot
10^{12}$ M$_\odot$.\\
The experimental data of the rotation velocity distribution 
in the Galactic disc, the so-called rotation curve (RC) of the MW,
 is adapted from the publication by
Sofue:2008, where different measurements with different
tracers are summarized, and the publication by Binney.
The RC shows a change of slope at a Galactocentric distance of about
10 kpc.

A $\chi^2$ fit of the density distribution is performed for cored 
and cuspy halo profiles. The experimental data allows no
differentiation between a cored and a cuspy DM halo since the density
distribution in the GC is dominated by the luminous matter. The local DM
density is obtained to be $\rho_{\mathrm{DM}, \odot} = 0.3 \pm 0.1$ GeV cm$^{-3}$.
For oblate haloes with $\epsilon_z = 0.7$ the local DM density is 
$\rho_{\mathrm{DM}, \odot} \approx 0.5$ GeV cm$^{-3}$ (Weber:2009). 
However, a poor description of the RC is found.\\

\begin{figure}
  \includegraphics[width=0.5\textwidth]{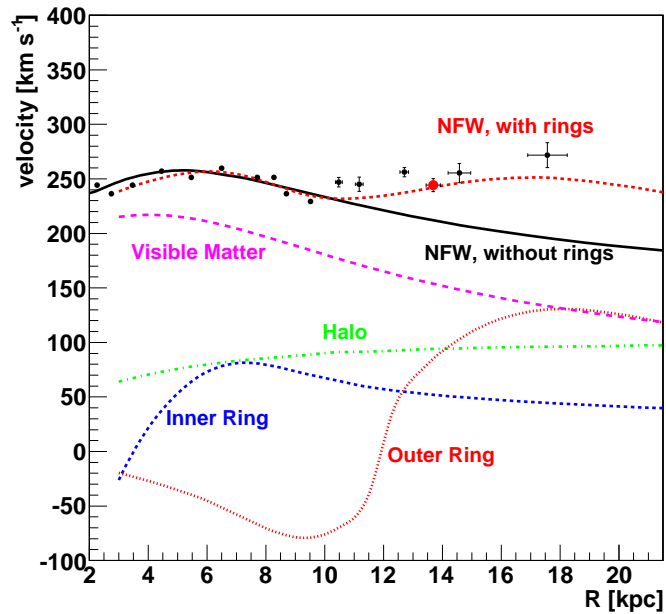}
  \caption{Rotation curve of the MW. The change of slope at a Galactocentric
           distance of about 10 kpc is clearly visible. A cuspy NFW profile
           in combination with two DM rings in the Galactic disc yields a 
           good description of the experimental data.}
  \label{fig:rc}
\end{figure}

In order to describe the change of slope in the RC and the
height of the interstellar gas distribution an additional DM component
in the Galactic disc is necessary. The gravitational influence of two 
doughnut-shaped DM rings (an inner ring at $r \approx 4$ kpc and a
outer ring at $r \approx 13$ kpc) leads to a change of slope in the 
RC at about 10 kpc as can be seen in Figure \ref{fig:rc}. 
This can be understand as follows.
The gravitational pull of the outer rings decelerate rotating objects
at the inner Galaxy and accelerates objects beyond the outer ring.
The outer ring also influences the gas flaring, i.e. the increase of the
gas layer with increasing distance from the Galactic centre. This is
simply to the decreasing gravitational potential. However, a ring of
DM in the outer Galaxy will increase locally the gravitational potential
and thus reduce the gas layer. This reduced gas flaring has indeed
be observed (Kalberla:2007), and provides independent evidence for a 
ring of DM. An enhancement of DM in the Galactic disc, so-called `dark
discs', have indeed been predicted by recent 
N-body simulations of the accretion of
satellite galaxies onto early galactic discs (Purcell:2009). Coplanar
tidal streams resulting from the disruption of the satellite galaxy only
feel the radial gradient of the gravitational potential of the Galaxy,
which leads to ringlike structures with a much longer lifetime than the
tidal streams in the halo.\\ 
The study of the stellar population of the outer disc (Newberg:2001) 
based on the observation of the fields at the Galactic anticentre with
the Sloan Digital Sky Survey (SDSS) showed a 
ring structure outside the main
spiral structure of the Galactic disc. This ring, which is unconnected to 
the spiral structure in the inner disc, is called outer ring or Monoceros
ring. An enhancement of stars along this ring was discovered in the Canis 
Major constellation (Bellazzini:2003, Bellazzini:2005) at 
Galactic longitudes around $240^\circ$. This overdensity was
interpreted as a dwarf galaxy, called Canis Major Dwarf, which could
be the progenitor of the tidal stream. The velocity dispersion of the
Canis Major stars is very low which further confirms their common 
origin (Martin:2005) and is not explainable with a warp of the
Galactic disc (Martin:2003).

In summary, the experimental constraints can be met by a cuspy and 
a cored profile. A differentiation between a cusp or a core in the GC
is not possible since the density distribution in the centre is 
dominated by the visible matter. The local DM density is determined
to be $\rho_{\mathrm{DM}, \odot} = 0.3 \pm 0.1$ GeV cm$^{-3}$ for
spherical haloes and $\rho_{\mathrm{DM}, \odot} \approx 0.7$ GeV cm$^{-3}$
for oblate haloes. The rotation curve and the height of the interstellar
gas distribution can be described assuming an additional ringlike DM component
in the Galactic disc which yields $\rho_{\mathrm{DM}, \odot} \le 1.0$ GeV 
cm$^{-3}$.\\
A large uncertainty of the local DM density is found which results
from the large uncertainty of the surface density of the visible matter.
Therefore, the accuracy of the estimation of the counting rate in direct
DM search experiments strongly depends on the determination of the surface density
of the visible matter.


{\bf References:}

\medskip

\begin{description}
\item{(Bellazzini:2003) Bellazzini, M. et al.,
Mon. Not. Roy. Astron. Soc. {\bf 354} (2004) 1263-1278 
[astro-ph/0311119].}
\vspace{-0.2cm}
\item{(Bellazzini:2005) Bellazzini, M. et al.,
Mon. Not. Roy. Astron. Soc. {\bf 366} (2006) 865-883
[astro-ph/0504494].}
\vspace{-0.2cm}
\item{(Binney) Binney, E. \& Merrifield, M. R.,
`Galactic astronomy', Princeton University Press (1998).}
\vspace{-0.2cm}
\item{(Bovy:2009) Bovy, J., Hogg, D. W. and Rix, H.-W.,
Astrophys. J. {\bf 704} (2009) 1704-1709
[astro-ph/0907.5423].}
\vspace{-0.2cm}
\item{(Cardone:2005) Cardone, V. F. and Sereno, M.,
(2005) [astro-ph/0501567].}
\vspace{-0.2cm}
\item{(Ghez:2008) Ghez, A. M. et al.,
(2008) [astro-ph/0808.2870].}
\vspace{-0.2cm}
\item{(Gillessen:2008) Gillessen, S. et al.,
Astrophys. J. {\bf 692} (2009) 1075-1109
[astro-ph/0810.4674].}
\vspace{-0.2cm}
\item{(Holmberg:2004) Holmberg, J. and Flynn, C.,
Mon. Not. Roy. Astron. Soc. {\bf 352} (2004) 440
[astro-ph/0405155].}
\vspace{-0.2cm}
\item{(Kalberla:2007) Kalberla, P. M. W., Dedes, L., Kerp, J. and Haud, U.,
(2007) [astro-ph/0704.3925].}
\vspace{-0.2cm}
\item{(Martin:2003) Martin, N. F. et al., Mon. Not. Roy. Astron. Soc. {\bf 348} (2004) 12 
[astro-ph/0311010].}
\vspace{-0.2cm}
\item{(Martin:2005) Martin, N. F. et al.,
Mon. Not. Roy. Astron. Soc. {\bf 362} (2005) 906-914
[astro-ph/0503705].}
\vspace{-0.2cm}
\item{(Naab:2005)
Naab, T. and Ostriker, J. P.,
Mon. Not. Roy. Astron. Soc. {\bf 366} (2006) 899-917
[astro-ph/0505594].}
\vspace{-0.2cm}
\item{(Newberg:2001) Newberg, H. J. et al.,
Astrophys. J. {\bf 569} (2002) 245-274 
[astro-ph/0111095].}
\vspace{-0.2cm}
\item{(Oort:1932) Oort, Jan H.,
Bulletin of the Astronomical Institutes of the Netherlands {\bf 6} (1932) 249}
\vspace{-0.2cm}
\item{(Purcell:2009) Purcell, C. W., Bullock, J. S. and Kaplinghat, M.,
Astrophys. J. {\bf 703} (2009) 2275-2284 
[astro-ph.GA/0906.5348].}
\vspace{-0.2cm}
\item{(Sofue:2008) Sofue, Y., Honma, M. and Omodaka, T., (2008) 
[astro-ph/0811.0859].}
\vspace{-0.2cm}
\item{(Sparke) Sparke, L. S. and Gallagher, J. S.,
``Galaxies in the Universe - An Introduction,''
Cambridge University Press (2007).}
\item{(Weber:2009) Weber, M. and de Boer, W.,
(2009) [astro-ph.CO/0910.4272].}
\vspace{-0.2cm}
\item{(Xue:2008) Xue, X. X. et al.,
Astrophys. J. {\bf 684} (2008) 1143-1158 
[astro-ph/0801.1232]}
\end{description}

\newpage

\section{Summary and Conclusions of the Workshop by H J de Vega and N G Sanchez}

Participants came from Europe, North and South America, Russia,  India, Korea.
Discussions and lectures were outstanding.
Inflection points in several current research lines emerged.

The participants and the programme represented the different communities doing
research on dark matter:

\begin{itemize}
\item{Observational astronomers}
\item{Computer simulators}
\item{Theoretical astrophysicists not doing simulations}
\item{Astroparticle theorists}
\end{itemize}

The {\bf hottest} subject of discussion was: 

What is the mass of the DM particle and what is its nature?

\medskip

A word about notation. We talk about 'keV scale DM particles'
instead of WDM which is less precise. Also,
we use the more precise name `wimp simulations'
indicating simulations with DM particles heavier than a GeV instead
of CDM. Since keV scale DM particles are non relativistic for $ z < 10^6 $
they also deserve the name of cold dark matter.

\medskip

Some conclusions are:

\begin{itemize}
\item{Facts and status of DM: Astrophysical observations
point to the existence of DM. Despite of that, proposals to
replace DM  by modifing  the laws of physics did appeared, however
notice that modifying gravity spoils the standard model of cosmology
and particle physics not providing an alternative.}
\item{The DM research appears to split in three sets:
(a) Particle physics DM: building models beyond the standard
model of particle physics, dedicated laboratory experiments,
annihilating DM. All concentrated on wimps.
(b) Astrophysical DM: astronomical observations, astrophysical models.
(c) Numerical cosmological simulations. The results of 
(b) and (c) do not agree with each other at small scales.}
\item{The DM domain of research is mature: there exists from
more than 20 years a large scientific community working in the subject;
there exists a large number of astronomical observations, there exists astrophysical
models which agree with the observations; all dedicated particle experiments of direct
search of wimps from more than twenty years gave nil results;
 many different groups perform N-body cosmological computer simulations and an important
number of conferences on DM and related subjects is held regularly. }
\item{Something is going wrong in the DM research.
What is going wrong and why?}
\item{Astronomical observations strongly indicate that
{\bf dark matter halos are cored till scales below 1 kpc}. 
More precisely, the measured cores {\bf are not} hidden cusps.}
\item{Numerical simulations with wimps (particles heavier than $ 1 $ GeV)
without {\bf and} with baryons yield cusped dark matter halos.
Adding baryons do not alleviate the problems of wimps simulations,
on the contrary adiabatic contraction increases the central density of cups
worsening the discrepancies with astronomical observations.}
\item{The observed galaxy surface density appears to be universal within $ \sim 10 \% $
with values around $ 100 \; M_{\odot}/{\rm pc}^2 $.}
\item{The results of numerical simulations must be confronted to observations.
The discrepancies of wimps simulations with the astronomical
observations at small scales $ \lesssim 100 $ kpc 
 {\bf keep growing and growing}: 
satellite problem (for example, only 1/3 of satellites predicted by wimps
simulations around our galaxy are observed), voids problem, peculiar velocities
problem (the observations show larger velocities than wimp simulations),
size problem (wimp simulations produce too small galaxies).}
\item{The use of keV scale DM particles in the simulations alleviate all the
above problems. For the core-cusp problem, setting
the velocity dispersion of keV scale DM particles seems beyond
the present resolution of computer simulations. Analytic work in the
linear approximation produces cored profiles for keV scale DM particles
and cusped profiles for wimps.}
\item{The features of electrons and positrons observed recently by Auger,
Pamela and HESS can all be explained as having their origin in the
explosions and winds of massive stars in the Milky Way.
All these observations 
of cosmic ray positrons and the like are due to normal astrophysical 
processes, and do not require special decaying or annihilation
of heavy DM particles.}
\item{None of the predictions of wimps simulations at small scales
(cusps, substructures, ...) have been observed.}
\item{Model-independent analysis of DM from phase-space density
and surface density observational data plus theoretical analysis
points to a DM particle mass in the keV scale.}
\item{As a conclusion, the dark matter particle candidates with large mass 
($ \sim 100$ GeV, the so called `wimps') became strongly disfavored,
while light (keV scale mass) 
dark matter are being increasingly favoured both from theory, numerical 
simulations and a wide set of astrophysical observations.}
\item{Many researchers (including several participants to this Workshop)
continue to work with heavy DM candidates
(mass $ \gtrsim 1 $ GeV) despite the {\bf growing} evidence that these
DM particles do not reproduce the small scale astronomical observations
($ \lesssim 100 $ kpc). Why? [The keV scale DM particles naturally produce the
observed small scale structure].  
The answer to this strategic question appears not to be strictly scientific but
it is anyway beyond the scope of these scientific conclusions. Such strategic
question was present in many animated discussions during the Workshop.}
\end{itemize}
It should be recalled that the connection between 
small scale structure features and the mass of the DM particle 
directly follows from the value of the free-streaming
length $ l_{fs} $ and is well known. Structures 
smaller than $ l_{fs} $ are erased by free-streaming.
DM particles with mass in the keV scale 
give $ l_{fs} \sim 100 $ kpc while 100 GeV DM particles produce an
extremely small $ l_{fs} \sim  0.1 $ pc. While $ l_{fs} \sim 100 $ kpc
is in nice agreement with the astronomical observations, a $ l_{fs} $ 
a million times smaller requires the existence of a host of
DM smaller scale structures till a distance of the size of the Oort's cloud
in the solar system. No structures of this type have ever been observed.

\bigskip

\begin{center}

All the Workshop lectures can be find at:

\bigskip

{\bf http://chalonge.obspm.fr/Programme\_CIAS2010.html}

\end{center}

\newpage
\section{Live Minutes of the Workshop by Peter Biermann}

\begin{center}

Peter Biermann$^{1,2,3,4,5}$

$^{1}$ MPI for Radioastronomy, Bonn, Germany; 
$^{2}$ Dept. of Phys., Karlsruher Institut f{\"u}r Technologie KIT, 
Germany, 
$^{3} $Dept. of Phys. \& Astr., Univ. of Alabama, Tuscaloosa, AL, USA; 
$^{4}$ Dept. of Phys., Univ. of Alabama at Huntsville, AL, USA; 
$^{5}$ Dept. of Phys. \& Astron., Univ. of Bonn, Germany ; 
\\
\end{center}

\vskip0.5cm

Dark matter has been detected 1933 (Zwicky) and basically behaves like a 
non-EM-interacting self-gravitating gas of particles. Observational 
arguments for it are the rotation curves of disk galaxies, the stability 
of thin disks, the hot gas in galaxies, galaxy groups and clusters of 
galaxies, as well as the Large Scale Structure of the Universe.

\subsection{Observations of galaxies}

This allows to go back to galaxy data to derive the key properties of 
the dark matter particle: Recent observations, Hogan \& Dalcanton (2000 
PRD, 2001 ApJ), Gilmore et al. (from 2006 MNRAS, 2007 ApJ, etc.), 
Strigari et al. (2008 Nature), clearly point to some common properties 
of galaxy cores.

\subsection{Our galaxy}

Markus Weber, KIT: Measuring the DM particle density: runs through the 
Jan Oort and Maarten Schmidt arguments, Holmberg \& Flynn 2004 
(0405155); uses the older data from Brunthaler; mentions a paper by 
Merrifield 1992 about the scale-height of gas in the GC region; gives 
indeed about 100 pc; seems to ignore the old Kahn \& Woltjer arguments 
on mass within a few hundred kpc; obtain $\rho_{DM} = 0.3 \pm 0.1 
GeV/cc$, in agreement with Salucci et al. (2010) - 1003.3101; using a 
different method they get $0.389 \, \pm \, 0.025$ GeV/cc (Bayesian 
Markov chain algorithm); DM rings can be produced by the infall of dwarf 
galaxies, and distort the rotation curves; he works with W. de Boer at 
Karlsruhe.

\subsection{Clusters of galaxies}

Alfonso Cavaliere, Rome: He argues that the intra-cluster medium 
constitutes the best plasma overall. ICP = Intra Cluster Plasma; uses 
entropy description to test plasma model for clusters, Molendi \& 
Pizzolato 2001; Pratt et al. 2010; entropy eroded by cooling, increased 
by shocks; Voit 2005, Arnaud et al. 2009; Fusco-Femiano, Lapi, Cavalieri 
2009; entropy floor and entropy ramp from central heating/cooling, and 
infall shock; results A2199 central ICP cold, A2597 cold, A1689 cold, 
A1656 hot, A2256 hot, A644 hot: in cold case DM extended, old, $z_t$ 
near 2; in hot case DM compact, young, $z_t$ near 0.5; deep mergers 
supply entropy, energy; M87 Forman et al. 2006, 15 kpc ring seen 
corresponding to explosion about $10^7$ years ago, with about $10^{59}$ 
ergs; Her A cluster McNamara \& Nulsen 2008, shocks at 200 kpc; Voit \& 
Donahue 2005; Ciotti \& Ostriker 2007; Tabor \& Binney 1998; Cavalieri 
\& Lapi 2008; Tozzi \& Norman 2001; Vikhlinin et al. 2009, Lau et al. 
2010, Molnar et al. 2010; Inogamov \& Sunyaev 2003; Reisprich et al. 
2009, Bautz et al. 2009, Geroeg et al. 2009,..; he confirms that AGN 
probably inject less energy than just major mergers -- I wonder whether 
that boils down to a tautology, since each major merger leads to feeding 
of a central BH.

\subsection{Small galaxies}

Matthew Walker, Cambridge, UK: Diemand, Kuhlen, Madau et al `Via Lactea 
II'; Belokurov et al. 2007; by definition star clusters do not need DM, 
dwarf galaxies do; globular clusters have a scale of order 10 pc, dwarf 
galaxies typically 100 pc; globular clusters single spike of star 
formation, dwarf galaxies continuous star formation; dwarf galaxies are 
entire dominated by DM, since they have about the same stellar 
luminosity; M/L ratio ranges up to hundreds in solar units, so they are 
the darkest stellar systems; best for DM detection; observes in the 
Magnesium triplet absorption lines; uses cross-correlation with a 
template spectrum; Walker et al. 2009; pressure supported systems are 
analyzed by velocity dispersion, gives flat curves; similar to normal 
galaxies show more matter than is seen directly; then analyzes data from 
dwarf galaxies, basically the details which Gerry Gilmore explained at 
previous Chalonge meetings; 
strongest constraint from using the mass within that 
radius, which contains half the light; see also Penarrubia et al. 2007, 
2008; Wolf et al. 2009, 2010; $ M(r_{half}) \, = \, \mu \; r_{half}  \; 
\sigma^2, \;  \mu \, = \, 580 \, M_{\odot} \, 
{\rm pc^{-1} \;  km^{-2} s^{2}} $; 
isothermal sphere ruled out; last author of Strigari et al. (2008), but 
now critical: claim then $M_{DM} (<300 pc) \, \simeq \, 10^{7} \, 
M_{\odot}$; for really small galaxies $M(<300 pc)$ not really well 
defined by data, used a math extrapolation; uses an extrapolation; the 
smaller galaxies may not even have a total of $10^{7}$ solar masses; but 
over several orders of magnitude the relationship in Strigari et al. 
(2008) is a good approximation, just not for the smallest galaxies; Marc 
Aaronson 1983 showed that dwarf galaxies are DM dominated, then based on 
three stars; Mateo 1993, 1008; ultrafaint dwarf galaxies deviate from a 
common dwarf spheroidals, showing a DM mass of order $10^{6}$ solar 
masses; universality is recovered with a quasi-universal mass profile, 
Walker et al. 2009; McGaugh et al. 2007 also gets $M_{DM} \, \sim \, 
r^2$; putting in the dwarf spheroidals , so again constant surface 
density: Walker et al. (2010) finds that dwarfs including M31 dwarfs do 
not give real universality, but close..; tidal stripping might just do 
it, but it would have to affect all dwarfs in M31; his final comment is 
that `details do matter'... recent paper by Penarrubia on tidal stripping 
from the disk shows that the M31 dwarfs can be understood.

\subsection{Galaxies}

Paolo Salucci, Trieste: Universality properties in galaxies and cored 
density profiles. Alludes to a paper with D. Sciama many years ago, in 
which they pointed out that DM is distributed very differently from 
baryonic matter. Uses inner rotation curves to derive limits on the 
presence of DM; for the biggest galaxies the rotation curves decline 
outwards, for the small ones they keep rising; introduces the Universal 
Rotation Curve URC; talks about M31, Corbelli 2010: all models fit 
except no DM; finds again, that small 
galaxies are completely dominated by DM, which big galaxies are 
dominated only in the outer parts; shows baryonic mass fct, and also 
halo mass fct; small galaxies have 100 times as much DM as baryonic 
matter, big galaxies only about ten times; inner part of rotation curve 
quite different from NFW; Bullock et al. 2001 MN 321, 559; Salucci et 
al. 2003 AA 409, 53; cusp vs core issue highlights a CDM crisis; quotes 
Klypin 2010; Gentile et al. 2004; cored halos best fits; DDO 47 good 
example, how NFW fails to fit in central parts; mentions van Eymeren; 
surface density in cores of small and big galaxies (elliptical?) always 
the same; but for stars this is not true, that surface density rises 
with mass; DM surface density rises; cored density means flat at center; 
key point is again that for big galaxies the rotation curves decline 
gently, and for small galaxies they rise gently - rotation curves are 
not flat.

\medskip

Gianfranco Gentile, Ghent: Surface densities and dark matter in 
galaxies, work with G. Gilmore, Salucci et al.: Pizzella et al. 2004; 
N3198, Begeman et al. 1991; CDM works very well on large scales, but 
has problems on galaxy scales, see Gentile et al. 2005; again, cored 
means constant density at center, NFW at center means spike at center, 
not seen; baryons would make CDM central densities even spikier by 
adiabatic contraction; de Blok \& Bosma 2002; Mashchenko et al. 2006; 
Governato et al. 2010; no consensus in community about how to use 
baryons in this problem; de Blok 2010: could it be that there are 
observational problems? Valenzuela et al. 2007 model of hot gas; Burkert 
(1995) $ \rho = \rho_0 r_c^3/(\{r + r_c\}
\{r^2 + r_c^2\}) $; Donato, Gentile et al. 2009; plotted central DM 
surface density all $141 \, M_{\odot} {\rm pc^{-2}}$ from dwarf 
spheroidal to large galaxies; this implies that the central acceleration 
generated by DM is a universal value of $ 3.2 \, 10^{-9} \, {\rm 
cm/s^2} $; see also Kormendy \& Freeman a long time ago; also the surface 
density of baryons (averaged over central scale) is constant, see GG et 
al. 2009 Nature; at the core radius of the DM the ratio of DM to baryons 
is the same everywhere, for every galaxy; at $r_0$ the gravitational 
acceleration of both baryons and DM is the same everywhere; see de Vega, 
Salucci, Sanchez 2010; HdV points out some analogy to BH entropy, as 
things go with surface; Donato, GG, et al. 2009 ! Norma S. keeps 
mentioning the Bekenstein bound on the thermodynamics of gravitational 
systems; $r_0$ is many kpc, so the mass of DM the correlation refers to 
is very large, much larger than the mass of the central BH.

\medskip

Janine Van Eymeren, Manchester/Duisburg-Essen: van Eymeren et al. 2009 
AA 505, 1 collaboration with various people, among them Dettmar 
(Bochum); she talks about the core/cusp discrepancy, discusses problems 
both with observations as with theory; only mergers between cored 
galaxies give a cored result; problems with simulations (Pedrosa et al. 
2009, Navarro et al. 2010); they looked at a sample of dwarf galaxies; 
HI synthesis observations; tilted ring analysis; quantify non-circular 
motions; decompose rotation curves into baryons and DM. So they looked 
at dwarf galaxies: N2366, N4861, N5408, I5152, 2 ESO galaxies; N2366 
gives a linear rotation curve in the central region; et al galaxies with 
lots of detail; they agree with the numbers of Donato et al. (2009), a 
paper which came out after their paper was accepted; result innermost 
slope close to zero, always core; `the measured cores are NOT hidden 
cusps'; review de Blok 2010.

\medskip

Chanda J. Jog, Bangalore: Determination of the density profile of DM 
halos of galaxies: she uses the observed HI distribution; starts with 
vertical shape of DM halos; in N3741 one can measure the HI disk to 40 
stellar disk length scales; Spitzer 1942; Narayan \& Jog AA 2002; 
Banerjee \& Jog 2008 ApJ; she finds an oblate halo for M31, with halo 
axis ratio 0.4; Bailin \& Steinmetz 2005, Bett et al. 2007 simulations 
give a range ob oblateness, 0.4 is at the most oblate end of what they 
find; Read et al. 2008: need baryons to simulate galaxies like M31; I 
bata (2008, unpubl.) confirms the oblateness from stellar stream; 
Banerjee, Matthews \& Jog 2010 New Astron. 15, 89; U7321 isothermal, 
spherical, high density halo: DM halo dominates all the way from $<$ 1 
kpc in this LSB galaxy; best fit requires high gas dispersion, which 
hinders star formation; Narayan, Saha \& Jog 2005 AA; Saha, Levine, Jog 
\& Blitz 2009 ApJ; Jog \& Combes 2009 Phys. Rep. Only cores are observed
in galaxies (never cusps). She does not find universal DM halo profiles 
contrary to many astronomical observations and cosmological simulations. 
This non-universality may be due to the fact that she uses in her fits 
several free parameters which reproduce non-universal structures.
In some galaxies the gas flares in the outer parts without any star 
formation to drive it.

\subsection{Simulations: Pure DM}

Anatoly Klypin, New Mexico SU: LCDM structure, Bolshoi simulation; talks 
a lot about halo concentration fct; distribution fct of halos above a 
circular velocity is a power-law, which strongly evolves with time; $n(> 
V) \, \sim \, V^{-3}$: Klypin et al. 2010; satellites follow DM for 0.2 
to 2 $R_{vir}$; Tissera et al. 2009; Gnedin et al. 2004; Duffy et al. 
2010; Tinker et al. 2008; Sheth \& Tormen ..; Kravtsov et al. 2004; 
Tasitsiomi et al. 2004; Conroy et al. 2006; Guo et al. 2009; 
Trujillo-Gomez et al. 2010; Spingob et al. 2007; Pizagno et al. 2007; 
Geha et al. 2006; luminosity vs circular velocity at 10 kpc: gives 
correct abundance of halos selected by circular velocity and places them 
in galaxies with the observed luminosity fct; Stark et al. 2009; Leroy 
et al. 2008; Williams et al. 2008?9; mass fct of halos at $z \, \simeq 
\, 6 \, - \, 10$ too low for reionization models; overabundance of 
dwarfs with $v_{circ} = 50 km/s$; Tikhonov \& Klypin 2008; Conclusion: 
standard explanation of overabundance of dwarf galaxies does not work: 
Most are predicted with $V_{circ} \, = \, 30 \, - \, 50 \, {\rm km/s}$, 
observed with $< \, 20 \, {\rm km/s}$: Big Problem - he calls himself 
`numerical observer'; he seems to get problems between LCDM simulation 
and observations exactly where WDM gets a free-streaming length.

\medskip

Andrea Lapi, Rome: Dark matter equilibria in galaxies and galaxy 
systems: Arguments based on simulations; density perturbation grows, 
detaches itself from expanding universe, then virializes; then merger 
tree, see Peebles 1993 e.g.; one finds a NFW profile $\rho \, \sim \, 
r^{-1} (1 + r)^{-2}$ with a cutoff; this is scale invariant; so the DM 
density profile of a galaxy like ours is approximately the same as that 
of a cluster of galaxies; Halo growth involves first a fast collapse 
including a few major violent mergers, second a slow accretion with the 
outskirts evolving from inside out by minor mergers and smooth mass 
addition; a universal powerlaw correlation $K = \sigma^2 \, \rho^{-2/3} 
\, \sim \, r^{\alpha}$; by analogy they call it DM entropy; $\alpha$ is 
1.25 to 1.3; density profiles are best fitted by Sersic-Einasto models: 
$\rho \sim r^{-\tau} \exp(-\frac{2 -r}{\eta} \times r^{\eta})$; this is 
also used for stelar mass profiles; see Fillmore \& Goldreich 1984, 
Austin et al. 2005, Lu et al. 2006; Lapi \& Cavaliere 2009a; see 
Bertschinger 1985; e.o.s. of DM is given by the entropy equation $\rho 
\, \sigma_r^2 \, \sim \, K \, \rho^{5/3} \sim r^{\alpha} \rho^{5/3}$; 
see Austin et al. 2005; one can show then that a typical 
$\alpha$-profile exists for every $\alpha \, < \, 35/27 \, = \, 
1.296..$; these solutions have an exponential cutoff at large radii, 
similar to NFW over intermediate radii, and flatter than NFW at small 
radii; see Lapi \& Cavalieri 2009b; see also Hansen \& Moore 2006; 
tested Abell1689; that cluster shows evidence for a higher transition 
redshift, and a result is a higher concentration; see Kormendy et al. 
2009 and Navarro et al. 2010; Prugniel \& Simien 1997; summary all from 
Lapi + ApJ 692, 174, 2009; ApJL 695, L125, 2009; AA 510, 90, 2010; ApJ 
698, 580, 2009; ApJ 705, 1019, 2009; AA 2010 in press. Klypin argues 
that this work is incompatible with simulations, disagreement; key point 
is use of a modified polytropic e.o.s., the entropy equation.

\medskip

Rainer Stiele, Uni HD: Cosmological bounds on DM self-interactions: 
starts again with small scale problem in LSSF; mentions again, that NFW 
fails in the center relative to data: Salucci et al. 2007; 
self-interacting DM heats center, also isotropization in central 
regions: Spergel \& Steinhardt 2000 PRL; Dave et al. 2001 ApJ; Yoshida 
et al. 2000 ApJ; Gnedin \& Ostriker 2001: galactic halos have to survive 
heating from hot cluster halos; Miralda-Escude et al. 2002 ApJ: 
ellipticity of cluster halos; ever more constraints; Randall et al. 2007 
ApJ bullet cluster; Clowe et al. 2006 ApJ; Clowe 2007; all give limits 
of order $\sigma_{DM}/m_{DM} \, < \, 0.1 \, {\rm cm^2/g}$; in cosmology 
period of self-interaction prior to radiation dominated era; scaling 
arguments lead to WDM; changes the early universe, no effect on MWBG, 
nucleosynthesis could be affected (better not); BBN then gives a 
constraint; could change the ${^4}He$ content; then $Helium{^4}$ 
abundance gives then a constraint on self-interaction: Steigman et al. 
2007 ARNPSc upper bound; Hui 2001 PRL 86; applying a similar approach to 
CDM gives $m_{CDM} \, < \, 45$ GeV ruled out if $\sigma \, > \, 
\sigma_{weak}$; then goes into structure formation; possible objects not 
compact, as pointed out by Andrea Maccio; then Norma S. and HdV say, 
that in their work there is self-interaction at a level 5 orders of 
magnitude below the present upper limits.

\subsubsection{DM with stars and gas. keV scale DM.}

Yehuda Hoffman, Hebrew U, Jerusalem, hoffman@huji.ac.il: Dark matter 
halos, with and without baryons: work with Isaac Rodriguez (Hebrew 
Univ.; his student), Emilio Romano-Diaz (these three one collaboration), 
Isaac Shlosman (Kentucky), Clayton Heller (Statesboro, GA, USA), Stefan 
Gottloeber (Potsdam), Gustavo Yepes (Madrid), Luis Martinez-Vaquero, 
Alexander Knebe, Steffen Knollmann. Last 5 CLUES collaboration, from 
Gottloeber. Two different kinds of halos, with his two different 
collaborations. The structure of DM halos is well known, but hardly 
understood analytically; no consensus on baryonic DM halos, no numerical 
convergence, no consensus on subgrid processes, results depend on 
implementation of subgrid processes; quotes the Blumenthal, Faber, 
Flores, Primack 1986 paper ApJ; $\{M_{DM}(r) + M_b(r)\} \, \times \, r $ 
adiabatic constant; Gnedin, Kravtsov, Klypin et al. 2004 ApJ; baryons 
pull DM into the center by cooling; al + Shlosman, Hoffman 2001 ApJ and 
2004 (second paper with J. Primack and F. Combes); dynamical friction 
transfers energy from one component to the other component; clumps need 
to be baryon rich; dynamical friction using Chandrasekhar 1943 formula; 
Anatoly Klypin calls him analytical simulator; Romano-Diaz, Shlosman, 
Heller, Hoffman (2008 - 2010); feedback from stellar winds and SNe, 
delayed cooling; very interesting phase space plots, for DM and baryons; 
shows some interesting plots of the evolution of a DM clump, stars and 
gas - I am a bit skeptical; at redshift z = 0 the baryons dominate at 
the center, but the DM density also enhanced at center compared to the 
non-baryon case; almost isothermal; Adiabatic contraction works, at all 
times the BDM (DM including the effects of baryons) density exceeds the 
PDM (pure DM excluding the effects of baryons); the excess of DM at 
center decreases with time; the dissipative gas makes the DM 
substructure more resilient against tidal forces, but the central galaxy 
potential gets deeper, so tidal forces increase; compared with the PDM 
the BDM subhalos die younger, lose more of their mass, lose more of 
their orbital energy, population is depleted faster; subhalo mass fct 
$M_{DM}^{-0.9}$; difference in central flattening between his 
Romano-Diaz et al collaboration, and Gottloeber et al collaboration; his 
plots are largely illegible; phase space density plots as fct(r) for DM 
only: gives radial powerlaw; in BDM case slightly shallower than in PDM 
case; possible contradiction between codes; dynamical friction time 
scale over dynamical time is scale free for subhalos; in massive halos 
($> \, 10^{12} \, M_{\odot}$) dynamical friction by subhalos flattens the 
central DM density cusp, in less massive halos does not flatten the 
inner density cusp (see also work by al + Ostriker 2009); conjecture 
that in more massive hosts the dynamical friction brings in more massive 
baryonic substructures, that puff up the central DM distribution; Paolo 
S. says, that M33 has a core in DM (no BH).

\medskip

Andrea Maccio, MPIA, HD: Dark matter at small scales, work with H.W. Rix 
et al.; will talk about LF of small satellites, from SDSS; Gnedin et al. 
2000 on reionization; started with N-body simulations, built merger 
tree; include tidal stripping, SFR, SN feedback, orbital evolution $->$ 
LF etc; Somerville et al. 2008, Kang et al. 2005, 2008, Morgana, Monaco 
et al. 2006, Gnedin 2000, Okamoto et al 2008; Maccio, Kang \& Moore 
2009, Maccio et al. 2010; Klypin et al. 1999; both SN feedback and 
reionization are able to change the LF; he claims to use the entire HRD 
for stellar evolution and mass loss; argues about reionization redshift, 
uses relatively small numbers like 7.5 to 11; for given DM mass a huge 
range of stellar mass; he can reproduce the Strigari et al. 2008 plot 
using CDM; quotes Maccio et al. 2008; Maccio \& Fontanot 2010, MN Lett; 
using WDM models for substructure give $m_{DM} \, > \, 1$ keV; he argues 
that Lyman-alpha and QSO lensing give $m_{DM} \, >$ 4 keV; he uses a 
thermal relic, while we already know that the relics have to be 
subthermal...; WDM also reproduces the Strigari et al. plot; he 
concludes that 3/4 of all satellite galaxies are dark, have only DM; 
heating of stellar streams (Oderkirchen et al. 2009); streams get 
substructure from dynamical friction with dark halos: the data look like 
they are disrupted occasionally, possibly proving the existence of dark 
halos; his small galaxies are too red; his final conclusion is that he 
cannot decide with respect to CDM vs WDM; from his point of view both 
are possible; big argument with Anatoly Klypin on small galaxies; Paolo 
Salucci tells that he and his team looked for satellites of large spiral 
galaxies, and detected 1/3 the predicted number; Yehuda Hoffman says 
that our Galaxy is embedded in two small filaments of the galaxy 
distribution towards Virgo.

\medskip

Gustavo Yepes, Madrid: How warm can dark matter be? Constraining the the 
mass of DM particles from the local universe: Collaboration CLUES 
`Constrained Local UniversE Simulation project': determine local flow 
field, extrapolate back, then use as initial conditions, and work 
forward again; since the nonlinear evolution strong, you learn 
conditions; they usually get complexes like Virgo and Fornax, but get a 
realistic local group 4 - 6 times out of 200 realizations; Hoffman et 
al. 2008 MN, Martinez-Vaquero et al. 0905.3134, Tikhonov et al. 2009 MN, 
Zavala et al., 2009 ApJ; says, that M31 has a linear orbit relative to 
our Galaxy; used then simulation to see, whether the radial orbit 
assumption gives reasonable masses (a la Kahn \& Woltjer 1959): this 
gives a typical factor 2 to 1/2; Tully-Fischer relation seems to work; 
efficient suppression of star formation by UV photoionization (Hoeft et 
al. 2006); substructure LF Metz 2007, Koposov 2008; Tikhonov \& Klypin 
2008; Tikhonov et al. 2009; study WDM simulations; in WDM models the 
density in the voids is slightly smaller than in corresponding CDM 
simulations; problem is with discreteness of numerical fluid at small 
scales; Wang \& White 2007; density profiles almost the same for CDM, 1 
keV WDM, and 3 keV WDM (Martinez-Vaquero 2010 PhD Univ. Aut. Madrid); 
did not use initial random velocities, so no core (central density 
flat); voids in local Universe, Tikhonov et al. 2009 MN: in WDM the void 
statistics work better, finding small galaxies; HI velocity function 
better fitted by WDM than by CDM models; their analysis points towards 1 
- 3 keV mass of the DM particle, depending on exact question; some 
arguments on various codes.

\subsection{DM with stars, gas, magnetic fields and cosmic rays}

We know and approximately understand due to Parker, that gas on one side 
and magnetic fields and cosmic rays on the other side are in pressure 
equilibrium in a Galactic disk. Similar constraints must govern the ISM 
in all galaxies, whether small or large. Cosmic rays can heat, but also 
wander around by diffusion and convection, can undergo adiabatic losses, 
but also gain energy from compression in shocks. Magnetic fields are 
probably initially produced by rotating stars in a combination of the 
battery and dynamo effects, injected into the ISM by stellar winds and 
supernova explosions, but then brought to full effect by a cosmic ray 
driven dynamo. The combined action of supernova explosions and cosmic 
rays then also drives a galactic wind. Clearly a full understanding of 
the interplay of gas and star formation requires the incorporation of 
the key physical elements of magnetic fields and cosmic rays.

\subsection{The Bullet Cluster \& $\Lambda$CDM.}

Jounghun Lee, Seoul NU: Bullet cluster is a challenge to LDCM! 1003.0939 
(ApJ in press). Bullet cluster one cluster 1.5 
peta-$M_{\odot}$, the other 1/10 of that; transverse velocity almost 
5000 km/s; probability of finding a bullet like cluster of 0.2 percent 
level (Hayashi \& White 2006); Farrar \& Rosen (2007) improved the 
likelihood by using better data, then likelihood $10^{-7}$; so bullet 
cluster quite unusual in LDCM; Springel \& Farrar (2007); therefore 
bullet cluster not so rare in LDCM, likelihood 0.07 (7 percent);
-- ; Mastropietro \& Burkert 2008: 
Her work has been to redetermine the likelihood, Crocce et al. 2010: 
MICE simulation; search for clusters of clusters with bullet like 
properties in simulation; probability then $ 10^{-10.5} $, at redshift 
$ z = 0.5 $ probability $ 10^{-8.5} $; he conclusion is: bullet cluster is 
incompatible with LCDM; Komatsu et al. 2001; Bradac et al. 2008 - other 
candidates to be similar to bullet cluster.

\subsection{Theory: Phase space and surface gravity constraints}

Hector de Vega, Paris: Galaxy properties from linear primordial 
fluctuations and keV dark matter from theory and observations. 
HdV talks about phase 
space redshift dependence; $ Q \, = \, \rho/\sigma^3 $ decreases by $ Z $: 
he estimates $ Z $ of order 40,000, simulations give less; in the 
estimates he uses $ 1 \, < \, Z \, < \, 10,000 $; allowing subthermal 
character gives slightly higher mass; Boyanovsky et al. PRD 77, 043.., 
2008, deVega \& Sanchez 2010 MN ; everything depends on the phase space 
density from dwarf galaxies from Gerry Gilmore 2007 and 2008; HdV says, 
that his result is independent to within a large factor of what the real 
phase space density is today, involves only the 1/4-power of the phase 
space density factor Z; HdV argues that this is independent of the 
cusp/core dispute; HdV uses the spatial average of the 
Q value, so it does not depend on structural details; shows primordial 
power spectrum today; for scales smaller than 100 kpc are cut off for 
keV particles, for larger scales the same as CDM; quotes universal 
properties like surface density, DM density profile, $M_{BH}/M_{halo}$; 
galaxy variables are related by universal empirical relations: one 
variable remains free; universal quantities may be attractors in 
dynamical evolution: $\rho(r) \, = \, \rho_0 \, F(r/r_0)$, $F(0) = 1, x 
= r/r_0$, $r_0$ core radius; e.g. Burkert $ 1/[(1+x)(1+x^2)] $; central 
density of galaxies $ 120 \, M_{\odot} \, {\rm pc^{-2}} \, = \, (17.6 
\,{\rm MeV})^3$ ; for $ 5 \, {\rm kpc} \, < \, r_0 \, < \, 100 \, {\rm 
kpc} $; compares with entropy arguments of BHs, where of course the 
corresponding energy is Planck scale; matching observations with 
formulae gives 2.6 keV (2.64.. for Bose, 2.69.. for Fermi particles); he 
concludes dark matter particle 1.6 keV $ < \, m_{DM} \, < \, 2 $ keV. WIMP 
mass of order 100 GeV disagree with observations by several orders of 
magnitude; he calls sterile neutrinos the simplest example; mixing angle 
$ 10^{-4} $; precise measurement of nucleus recoil in tritium beta decay; 
summarizes a) phase space density, b) proper galaxy density profiles, c) 
peculiar velocities in clusters,...;  arguments on supersymmetry.
I reminded the audience of the advantage of early star 
formation (PLB \& AK 2006 PRL); HdV and Norma S both argue about 
supersymmetry, both say, that supersymmetry by itself does NOT give a 
heavy DM particle, you also need another symmetry (R-parity), Norma S 
says, that supersymmetry should appear at much higher energy scale in 
the universe; HdV emphasizes that with the two mass scales, right-handed 
neutrino and left-handed neutrino they just give the right abundance 
from the mixing angle.

\medskip

Norma Sanchez, Paris: Galaxy properties, keV scale DM from theory and 
observations, and the power of linear approximations: 1) DM exists, 2) 
Astrophysical observations point to existence of DM, 3) no WIMP search 
successful, 4) Proposals keep coming to make DM disappear, changing the 
laws of physics... lecture A) the mass of the DM particle, 
B) Boltzmann-Vlasov eq, 
C) universal properties of galaxies. Various papers from 2008 through 
2010, MNRAS, ApJ, etc, and astro-ph; collaborators 
Boyanovsky, de Vega, Salucci, ..; 
central density profiles, cores rather than cusps (at very center 
density law flat; DM energy density, DM velocity dispersion, DM phase 
space density together give DM particle mass and decoupling temperature; 
allowed range for velocity dispersion, range of radial scale, gives 
phase space density to within a factor of 10; $Q = \rho/\sigma^3$ 
decreases by nonlinear gravitational interactions (Lynden-Bell, 
Tremaine, Henon 1986);free-streaming length gives $ 450 \, M_{\odot} \, < 
\, M_{z} (1+z)^{-3/2} \, < \, 4.5 \, 10^{6} \, M_{\odot} $; this length 
decreases with increasing mass of the particle 8 kpc for 1 keV, and 2 
kpc at several keV; decoupling temperature around 100 GeV; DM 
annihilation cross-section about $ 10^{-9} \, {\rm GeV^{-2}} $, about 
$10^{5}$ below other limits; for WIMPs the free streaming length is of 
order 100 AU; so WIMPS strongly disfavored; she pushes universality, 
'the first things to understand'; constant central surface density $120 
\, M_{\odot} \, {\rm pc^{-2}}$ over core radius 5 to 100 kpc: Gentile et 
al. 2009, Donato et al. 2009; gets at the end a universal density 
profile for galaxies, with just the length scale as parameter; radial 
scaling Walker et al. 2009 (observations), Vass et al. 2009 
(simulations); she says that the agreement between linear theory and 
observations is remarkable; she gives a range for the DM particle of 1.6 
keV $ < \, m_{DM} \, < \, 2 $ keV; reminds of the review de Blok 2010; 
Hoffman et al. 2007; Avila-Reese et al. 2000, Goetz \& Sommer-Larsen 
2002; Tikhonov et al. 2009; Kashlinsky et al. 2008; Watkins et al. 2009; 
Lee \& Komatsu 2010; Ryan Joung et al. 2009, Holz \& Perlmutter 2010; 
Blasi, Serpico 2009 (pulsar wind nebulae); Maccio 
wonders why heavy particles give different 
velocities in entire galaxy situations, asking about Lee \& Komatsu 
2010; Norma quotes Boyanovsky, HdV \& Norma Sanchez 2004 and say, 
that they 
generalize Gunn \& Tremaine bound, a paper much earlier; so step by step 
evolution Gunn et al., Hogan et al., now Boyanovsky et al.. 
Paolo S then says, that 
he can compute phase space evolution; Norma Sanchez mentions the issue whether baryons 
could also solve all the short scale problems using wimps - and argues 
that this will not be possible; argues that universal quantities cannot 
be changed by a small fraction of baryons; but acknowledges that the 
simulations have not all been done yet; now Maccio argues that 
his calculations can account for substructure using wimps and baryons - 
the point is that in his approach many small units are invisible; 
N Sanchez argues that this implies many different recipes to solve the different 
problems, she prefers a unique framework not different recipes 
and by Occam's razor keV particles are a simpler solution. 
One would really need keV DM simulations with baryons.

\subsection{Black holes, pulsar kick}

The transformation of a left-handed active neutrino into a right-handed 
sterile neutrino can give a pulsar a kick (Kusenko 2004 IJMPD), possibly 
explaining all the high linear velocities found for a fraction of all 
pulsars.

\subsection{Decisive Observations to find the DM particle}

\subsubsection{Very high precision experiments?}

Basically all extremely high precision experiments have the chance to 
discover basic physics: examples are experiments done by Quack (Quack et 
al. Ann. Rev. Phys. Chem vol. 59, 2008) , H{\"a}nsch (Science 319, 1808, 
2008) , Marcaide \& Irwin Shapiro (IAU Sympos. 110, p.361, 1984).

The first is molecular spectroscopy, the second one uses frequency 
combs, and the third one micro-arcsecond angular precision astrometry 
across the universe.

These experimental methods may lead to discover the keV scale
DM particle.

\subsubsection{X-ray emission line}

In the case of a right-handed sterile neutrino there should be a 
weak decay line 
at a photon energy of just half the mass (e.g. Loewenstein et al. 2009 
ApJ; Boyarsky et al. arXiv:1001.0644). This would be proof, if detected 
with the right spatial profile.

\subsubsection{Very early star formation}

The perhaps most exciting prediction from the concept that DM is a keV 
right handed  sterile neutrino is that massive star formation, black hole 
formation, and perhaps even supermassive black hole formation start in 
the redshift range 50 to 1000 (Biermann \& Kusenko 2006 PRL).

\subsection{Heavy dark matter particle decay?}

 From particle physics Supersymmetry suggests with an elegant argument 
that there should be a lightest supersymmetric particle, which is a dark 
matter candidate, possibly visible via decay in odd properties of 
energetic particles and photons:

\medskip

Gabrijela Zaharijas, Gif-sur-Yvette: DM constraints from Fermi-LAT 
diffuse observations: on behalf of the Fermi collaboration; launched 
exactly 2 years ago, June 11, 2008; she uses the LambdaCDM model with 
WIMPs; the decay of these postulated WIMPs then gives gamma rays; 
Bergstrom et al. 2009 PRL gives $E^{-3}$ CR-e spectrum; Gustafsson et 
al. PRL 2007; Abdo et al. 2010 PRL diffuse signal; considers enhancement 
of diffuse flux due to formation of gravitational structures: 
extrapolation to below the spatial resolution of the simulation, Bullock 
et al. 2001, Zavala et al. MN 405, 593 (?); Ullio et al. PRD 2002; Eke 
et al. 2001 ApJ; Abdo et al. JCAP 2010; Primack, Gilmore, Somerville 
0811.3230; Gilmore et al. 2009, 0905.1144; Stecker et al. 0510449; star 
forming galaxies could make up most of the extragalactic signal, AGN 
only $<$ 30 percent; Fields et al. 1003.3647; as with increased 
sensitivity more and more point sources are resolved, the diffuse 
background goes down, and gets more constraining; comment from the 
audience, that EGRET saw a much higher diffuse flux, and then it was 
also claimed that AGN could account for 30 percent: response was that 
now we have better data on AGN, and their flux distribution;

Observations have discovered i) an upturn in the CR-positron fraction 
(Pamela: Adriani et al. 2009 Nature), ii) an upturn in the CR-electron 
spectrum (ATIC: Chang et al. 2008 Nature; Fermi: Aharonian et al. 2009 
AA), iii) a flat radio emission component near the Galactic Center (WMAP 
haze: Dobler \& Finkbeiner 2008 ApJ), iv) a corresponding IC component 
in gamma rays (Fermi haze: Dobler et al. 2009 arXiv), v) the 511 keV 
annihilation line also near the Galactic Center (Integral: 
Weidenspointner et al. 2008 NewAR), and most recently, vi) an upturn in 
the CR-spectra of all elements from Helium (CREAM: Ahn et al. 2009 ApJ, 
2010 ApJL).

All these features can be quantitatively explained with the action of 
cosmic rays accelerated in the magnetic winds of very massive stars, 
when they explode (Biermann et al. 2009 PRL, 2010 ApJL), based on 
well-defined predictions from 1993 (Biermann 1993 AA, Biermann \& 
Cassinelli 1993 AA, Biermann \& Strom 1993 AA, Stanev et al 1993 AA). 
While the leptonic part of these observations may be explainable with 
pulsars and their winds (talk by P. Serpico), the hadronic part clearly 
needs very massive stars, such as Wolf-Rayet stars, their winds and 
their explosions. What the cosmic ray work (Biermann et al., from 1993 
through 2010) shows, that allowing for the magnetic field topology of 
Wolf Rayet star winds (see, e.g. Parker 1958 ApJ), both the leptonic and 
the hadronic part get readily and quantitatively explained, so by 
Occam's razor the Wolf-Rayet star wind proposal is much simpler.

\medskip

Pasquale Serpico, CERN: Pamela/Fermi CR lepton data... he mentions 
pulsar wind nebulae, SNRs; Nature 458, 607, 2009; basically follows the 
Strong \& Moskalenko line; PRL 102, 181101 (2009); 0909.4548 Di Bernardo 
et al.; Fermi LAT PRL 103, 251101 (2009); Stawarz et al. 0908.1904; D. 
Grasso et al. 0905.0636; Shaviv et al. PRL 103, 111302 (2009); Kobayashi 
et al. ApJ 601, 340 (2004); uses a lot of standard arguments; talks a 
lot about pulsar power; rotational energy of pulsars about two orders of 
magnitude larger than need to explain Pamela excess; question how to 
covert a large fraction of Poynting flux into nonthermal particles 
without any visible thermal flux; Amato \& Arons ApJ 653, 325 (2006); 
Hoshino \& Arons Phys of Fluids B 3, 818 (1991); he pushes pulsar wind 
nebulae as a source for extra CR-positrons and extra CR-electrons. In 
questioning he argues that with WR-star model you need special 
parameters, just as with pulsar wind model; I suggested that the polar 
cap component has now been confirmed through the fit to the CREAM data; 
then he separated hadronic from leptonic CR data.

\medskip

In summary there are convincing arguments, that all these observations 
of cosmic ray positrons and the like are due to normal astrophysical 
processes, and do not require a special heavy particle to decay.

\subsection{Conclusion}

A right-handed  sterile neutrino is a candidate to be this DM particle 
(e.g. Kusenko \& Segre 1997 PLB; Fuller et al. 2003 PRD; 
Kusenko 2004 IJMP; for a 
review see Kusenko 2009 PhysRep; Biermann \& Kusenko 2006 PRL; Stasielak 
et al. 2007 ApJ; Loewenstein et al. 2009 ApJ): This particle has the 
advantage to allow star formation very early, near redshift 80, and so 
also allows the formation of supermassive black holes, possibly formed 
out of agglomerating massive stars. Black holes in turn also merge, but 
in this manner start their mergers at masses of a few million solar 
masses. This readily explains the supermassive black hole mass function. 
The corresponding gravitational waves are not constrained by any 
existing limit, and could have given a substantial energy contribution 
at high redshift.

Our conclusion is that a right-handed  sterile neutrino of a mass of a 
few keV is the most interesting candidate to constitute dark matter.

\subsection{Acknowledgements}

PLB would like to thank G. Bisnovatyi-Kogan, J. Bl{\"u}mer, R. Engel, 
T.K. Gaisser, L. Gergely, G. Gilmore, A. Heger, G.P. Isar, P. Joshi, 
K.H. Kampert, Gopal-Krishna, A. Kusenko, N. Langer, M. Loewenstein, I.C. 
Mari\c{s}, S. Moiseenko, B. Nath, G. Pavalas, E. Salpeter, N. Sanchez, 
R. Sina, J. Stasielak, V. de Souza, H. de Vega, P. Wiita, and many 
others for discussion of these topics.

\subsection{References}

\begin{description} 
\item[1] Adriani, O., et al. (Pamela Coll.), \Nature {\bf 458}, 607 - 
609 (2009); arXiv 0810.4995
\vspace{-0.2cm}
\item[2] Aharonian, F., et al., (H.E.S.S.-Coll.), \AA {\bf 508}, 561 - 
564 (2009); arXiv:0905.0105
\vspace{-0.2cm}
\item[3] Ahn, H.S. et al. (CREAM-Coll.), \ApJ {\bf 707}, 593 - 603 
(2009); arXiv:0911.1889
\vspace{-0.2cm}
\item[4] Ahn, H.S. et al. (CREAM-Coll.), \ApJL {\bf 714}, L89 - L93 
(2010); arXiv:1004.1123
\vspace{-0.2cm}
\item[5] Biermann, P.L., \AA {\bf 271}, 649 (1993) - paper CR-I; 
astro-ph/9301008
\vspace{-0.2cm}
\item[6] Biermann, P.L., \& Cassinelli, J.P., \AA {\bf 277}, 691 
(1993), paper CR-II;astro-ph/9305003.
\vspace{-0.2cm}
\item[7] Biermann, P.L., \& Strom, R.G., \AA {\bf 275}, 659 (1993) - 
paper CR-III;astro-ph/9303013
\vspace{-0.2cm}
\item[8] Biermann, P.L., 23rd ICRC, in Proc. `Invited, Rapporteur and 
Highlight papers'; Eds. D. A. Leahy et al., World Scientific, Singapore, 
p. 45 (1994)
\vspace{-0.2cm}
\item[9] Biermann, P. L., Becker, J. K., Meli, A., Rhode, W., Seo, 
E.-S., \& Stanev, T., \PRL {\bf 103}, 061101 (2009); arXiv:0903.4048
\vspace{-0.2cm}
\item[10] Biermann, P.L., Becker, J.K., Caceres, G., Meli, A., Seo, 
E.-S., \& Stanev, T., \ApJL {\bf 710}, L53 - L57 (2010); arXiv:0910.1197
\vspace{-0.2cm}
\item[11] Bisnovatyi-Kogan, G. S., {\it Astron. Zh.} {\bf 47}, 813 (1970)
\vspace{-0.2cm}
\item[12] Bisnovatyi-Kogan, G. S., Moiseenko, S. G., {\it Chinese J. of 
Astron. \& Astroph. Suppl.} {\bf 8}, 330 - 340 (2008)
\vspace{-0.2cm}
\item[13] Chang, J., et al. \Nature {\bf 456}, 362 (2008)
\vspace{-0.2cm}
\item[14] Dobler, G., Finkbeiner, D.P., \ApJ {\bf 680}, 1222 - 1234 
(2008); arXiv:0712.1038
\vspace{-0.2cm}
\item[15] Dobler, G., Finkbeiner, D. P., Cholis, I., Slatyer, T. R., 
Weiner, N., eprint arXiv:0910.4583 (2009)
\vspace{-0.2cm}
\item[16] Gopal-Krishna, Peter L. Biermann, Vitor de Souza, Paul J. Wiita,
in press \ApJL (2010); arXiv:
\vspace{-0.2cm}
\item[17] Schlickeiser, R., Ruppel, J., \NJPh {\bf 12}, 033044 (2010); 
arXiv:0908.2183
\vspace{-0.2cm}
\item[18] Stanev, T., Biermann, P.L. \& Gaisser, T.K., \AA {\bf 274}, 
902 (1993) - paper CR-IV; astro-ph/9303006
\vspace{-0.4cm}
\item[19] Stawarz, L., Petrosian, V., \& Blandford, R.D., \ApJ {\bf 
710}, 236 - 247 (2010); arXiv:0908.1094
\vspace{-0.2cm}
\item[20] Weidenspointner, G., et al. \NewAR {\bf 52}, 454 - 456 (2008)
\end{description}

\newpage

\section{List of Participants}

LAST NAME FIRST NAME, INSTITUTION, CITY, COUNTRY.

\bigskip

AMES Susan, Oxford Astrophysics, Oxford, ENGLAND.

\medskip

BIERMANN  Peter, MPIfR Bonn, Germany and Univ. Alabama, USA.

\medskip

CASARINI Luciano, Universita' Milano Bicocca - INFN Milano ITALY.  

\medskip

CAVALIERE Alfonso.  Dipt. Fisica/Astrofisica, Univ. Roma 2, Tor Vergata, Rome, ITALY.  

\medskip

CHERNENKO Anton, IKI, Moscow, RUSSIA.  

\medskip

CNUDDE     Sylvain, LESIA Observatoire de Paris,     Meudon, FRANCE.  

\medskip

DE VEGA  Hector J.,  LPTHE, Univ Pierre \& Marie Curie \& CNRS, Paris, FRANCE .

\medskip

ECKART Andreas I., Physikalisches Institute, Cologne, GERMANY. 

\medskip

EMRE Kahya, Friedrich-Schiller-Universit\"at, Jena, GERMANY. 

\medskip

FISCHER Oliver, Albert Ludwigs Universit\"at Freiburg, Freiburg GERMANY  

\medskip

FONTIJN  Leen, Lightning Flash Consultancy, Lansingerland, THE NETHERLANDS.  

\medskip

GAVAZZI Raphael, Institut d'Astrophysique de Paris, Paris, FRANCE.  

\medskip

GENTILE Gianfranco,  Mathematical Physics \& Astronomy, University of Ghent, Ghent, BELGIUM  

\medskip

GUPTA Rajiv Guru Nanak Dev, University Amritsar, INDIA.  

\medskip

HASHIM Norsiah, University of Malaya, Kuala Lumpur, MALAYSIA.  

\medskip

HOFFMAN Yehuda, Racah Institute of Physics, Hebrew Univ of Jerusalem, ISRAEL. 

\medskip

JACHOLKOWSKA Agnieszka, LPNHE, Paris, FRANCE. 

\medskip

JOG Chanda, Indian Institute of Science, Bangalore, INDIA.  

\medskip

KHACHATRYAN Suren, American University of Armenia, Yerevan, ARMENIA.  

\medskip

KHADEKAR Goverdhan RTM Nagpur University Nagpur INDIA  

\medskip

KLYPIN  Anatoly Dept. of Astronomy, New Mexico State University Las Cruces USA  

\medskip

KONTUSH Anatol INSERM Paris FRANCE  

\medskip

KRUSBERG Zosia University of Chicago Chicago USA   

\medskip

LALOUM  Maurice CNRS/IN2P3/LPNHE Paris FRANCE

\medskip

LA VACCA  Giuseppe  Univ. Milano-Bicocca  Milano ITALY  

\medskip

LAPI Andrea  Dipt Fisica/Astrofisica, Univ Roma 2 Tor Vergata Rome ITALY  

\medskip

LEE  Jounghun  Dep. of Physics \& Astronomy, Seoul National University Seoul SOUTH KOREA   

\medskip

LEPETIT PIERRE retraite Bretigny sur or FRANCE  

\medskip

LETOURNEUR Nicole Observatoire de Paris LESIA Meudon  Meudon FRANCE  

\medskip

LI Nan National Astronomical Observatories, Chinese Academy of Sciences Beijing CHINA  

\medskip

LOPES Dominique  LERMA Observatoire de Paris Meudon FRANCE  

\medskip

MACCIO Andrea V.  Max Planck Institut fur Astronomie, Heidelberg Heidelberg GERMANY  

\medskip

MURANTE Giuseppe INAF- Osservatorio di Torino Torino ITALY  

\medskip

NAUMANN Christopher LPNHE-IN2P3 Paris FRANCE  

\medskip

PEIRANI Sebastien Institut d'Astrophysique de Paris Paris FRANCE  

\medskip

PILIPENKO Sergey Lebedev Physical Institute Moscow RUSSIA  

\medskip

PILO Luigi University of L'Aquila L'Aquila ITALY  

\medskip

RAMON MEDRANO Marina Universidad Complutense.  Madrid SPAIN  

\medskip

RAZZAQUE MD Abdur Weimi Bangladesh LTD Kushtia BANGLADESH   

\medskip

SALUCCI  Paolo  SISSA, Astrophysics Group Trieste ITALY  

\medskip

SANCHEZ Norma G. Observatoire de Paris LERMA \& CNRS Paris FRANCE  

\medskip

SEVELLEC Aurelie Observatoire de Paris LESIA Meudon  Meudon FRANCE  

\medskip

SMOOT  George  F. Univ Paris Diderot \& Univ California Berkeley  Paris  FRANCE  

\medskip

STIELE  Rainer  Institute for Theoretical Physics, Heidelberg Univ Heidelberg GERMANY  

\medskip

TRONCOSO Iribar Paulina Alejandra, Osservatorio Astronomico di Roma, Rome, ITALY. 

\medskip

VALLS-GABAUD David, GEPI - Observatoire de Paris, Meudon, FRANCE.

\medskip

VAN EYMEREN  Janine,  Univ of Manchester UK \& Duisburg-Essen, Germany.

\medskip

WALKER  Matthew G.  Institute of Astronomy, University of Cambridge,
Cambridge, UNITED KINGDOM  

\medskip

WEBER  Markus,  Inst fur Experimentelle Kemphysik, KIT Karlsruher, 
Karlsruhe, GERMANY. 

\medskip

YEPES Gustavo, Grupo Astrofisica, Univ Autonoma de Madrid, 
Cantoblanco, Madrid, SPAIN   

\medskip

ZAHARIJAS  Gabrijela, IPhT/CEA Saclay,  FRANCE.

\medskip

ZIDANI Djilali, LERMA Observatoire de Paris - CNRS, Paris, FRANCE.

\begin{figure}[htbp]
\epsfig{file=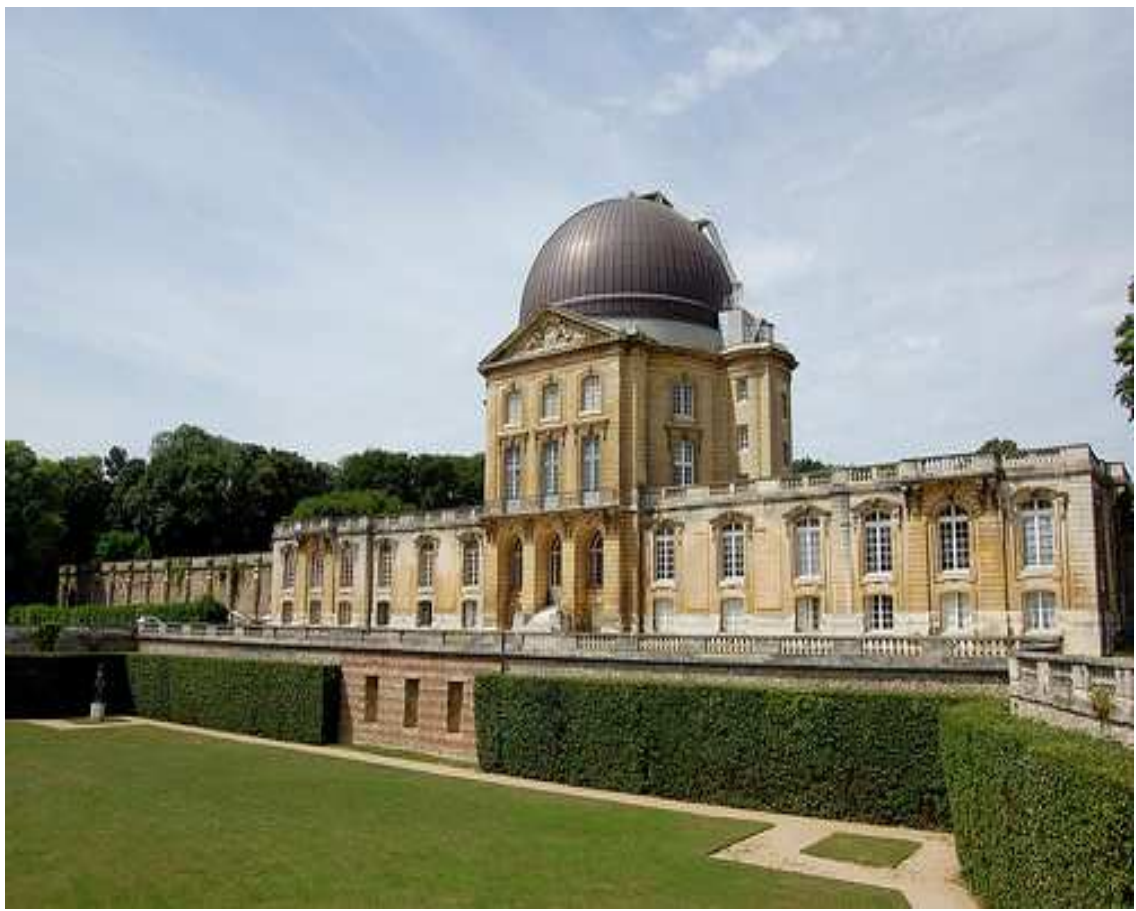,width=15cm,height=12cm}
\caption{The Meudon Ch\^ateau}
\end{figure}

\end{document}